\documentclass[twocolumn]{aastex61}
\usepackage{longtable}

\submitjournal{ApJ}

\shorttitle{A Calibrator Sample for the M dwarf Abundances}
\shortauthors{Souto et al.}

\begin{document}

\title{Detailed Chemical Abundances for a Benchmark Sample of M Dwarfs from the APOGEE Survey}
\correspondingauthor{Diogo Souto}
\email{diogosouto@academico.ufs.br}

\author[0000-0002-7883-5425]{Diogo Souto}
\affiliation{Departamento de F\'isica, Universidade Federal de Sergipe, Av. Marechal Rondon, S/N, 49000-000 S\~ao Crist\'ov\~ao, SE, Brazil}

\author[0000-0001-6476-0576]{Katia Cunha}
\affiliation{Steward Observatory, University of Arizona, 933 North Cherry Avenue, Tucson, AZ 85721-0065, USA}
\affiliation{Observat\'orio Nacional/MCTIC, R. Gen. Jos\'e Cristino, 77,  20921-400, Rio de Janeiro, Brazil}

\author[0000-0002-0134-2024]{Verne V. Smith}
\affiliation{NOIRlab, 950 North Cherry Avenue, Tucson, AZ 85719, USA}

\author{C. Allende Prieto}
\affiliation{Instituto de Astrof\'isica de Canarias, E-38205 La Laguna, Tenerife, Spain}
\affiliation{Departamento de Astrof\'isica, Universidad de La Laguna, E-38206 La Laguna, Tenerife, Spain}


\author{Kevin Covey}
\affiliation{Department of Physics \& Astronomy, Western Washington University, Bellingham, WA, 98225, USA}

\author[0000-0002-1693-2721]{D. A. Garc\'ia-Hern\'andez}
\affiliation{Instituto de Astrof\'isica de Canarias, E-38205 La Laguna, Tenerife, Spain}
\affiliation{Departamento de Astrof\'isica, Universidad de La Laguna, E-38206 La Laguna, Tenerife, Spain}

\author[0000-0002-9771-9622]{Jon A. Holtzman}
\affiliation{New Mexico State University, Las Cruces, NM 88003, USA}

\author[0000-0002-4912-8609]{Henrik J\"onsson}
\affiliation{Materials Science and Applied Mathematics, Malm\"o University, SE-205 06 Malm\"o, Sweden}

\author{Suvrath Mahadevan}
\affiliation{Department of Astronomy \& Astrophysics, Pennsylvania State, 525 Davey Lab, University Park, PA 16802, USA}
\affiliation{Center for Exoplanets \& Habitable Worlds, Pennsylvania State, 525 Davey Lab, University Park, PA 16802, USA}

\author{Steven R. Majewski}
\affiliation{Department of Astronomy, University of Virginia, Charlottesville, VA 22904-4325, USA}

\author{Thomas Masseron}
\affiliation{Instituto de Astrof\'isica de Canarias, E-38205 La Laguna, Tenerife, Spain}
\affiliation{Departamento de Astrof\'isica, Universidad de La Laguna, E-38206 La Laguna, Tenerife, Spain}

\author{Marc Pinsonneault}
\affiliation{Department of Astronomy, The Ohio State University, Columbus, OH 43210, USA}

\author{Donald P. Schneider}
\affiliation{Department of Astronomy \& Astrophysics, Pennsylvania State, 525 Davey Lab, University Park, PA 16802, USA}

\author{Matthew Shetrone}
\affiliation{University of Texas at Austin, McDonald Observatory, Fort Davis, TX 79734, USA}

\author[0000-0002-3481-9052]{Keivan G. Stassun}
\affiliation{Department of Physics and Astronomy, Vanderbilt University, 6301 Stevenson Center Ln., Nashville, TN 37235, USA}

\author{Ryan Terrien}
\affiliation{Department of Physics \& Astronomy, Carleton College, Northfield MN, 55057, USA}

\author{Olga Zamora}
\affiliation{Instituto de Astrof\'isica de Canarias, E-38205 La Laguna, Tenerife, Spain}
\affiliation{Departamento de Astrof\'isica, Universidad de La Laguna, E-38206 La Laguna, Tenerife, Spain}

\author{Guy S. Stringfellow}
\affiliation{Center for Astrophysics and Space Astronomy, University of Colorado, Campus Box 389, Boulder, CO 80309-0389}

\author{Richard R. Lane}
\affiliation{Centro de Investigación en Astronomía, Universidad Bernardo O'Higgins, Avenida Viel 1497, Santiago, Chile}

\author[0000-0003-4752-4365]{Christian Nitschelm}
\affiliation{Centro de Astronom{\'i}a (CITEVA), Universidad de Antofagasta, Avenida Angamos 601, Antofagasta 1270300, Chile}

\author[0000-0002-0149-1302]{B\'arbara Rojas-Ayala}
\affiliation{Instituto de Alta Investigaci\'on, Universidad de Tarapac\'a, Casilla 7D, Arica, Chile.}


\begin{abstract}
Individual chemical abundances for fourteen elements (C, O, Na, Mg, Al, Si, K, Ca, Ti, V, Cr, Mn, Fe, and Ni) are derived for a sample of M-dwarfs using high-resolution near-infrared $H$-band spectra from the SDSS-IV/APOGEE survey. 
The quantitative analysis included synthetic spectra computed with 1-D LTE plane-parallel MARCS models using the APOGEE DR17 line list to determine chemical abundances. 
The sample consists of eleven M-dwarfs in binary systems with warmer FGK-dwarf primaries and ten measured interferometric angular diameters. 
To minimize atomic diffusion effects, [X/Fe] ratios are used to compare M-dwarfs in binary systems and literature results for their warmer primary stars, indicating good agreement ($<$0.08 dex) for all studied elements. The mean abundance differences in Primaries-this work M-dwarfs is -0.05$\pm$0.03 dex. It indicates that M-dwarfs in binary systems are a reliable way to calibrate empirical relationships.
A comparison with abundance, effective temperature, and surface gravity results from the ASPCAP pipeline (DR16) finds a systematic offset of [M/H], $T_{\rm eff}$, log$g$ = +0.21 dex, -50 K, and 0.30 dex, respectively, although ASPCAP [X/Fe] ratios are generally consistent with this study.
The metallicities of the M dwarfs cover the range of [Fe/H] = -0.9 to +0.4 and are used to investigate Galactic chemical evolution via trends of [X/Fe] as a function of [Fe/H]. 
The behavior of the various elemental abundances [X/Fe] versus [Fe/H] agrees well with the corresponding trends derived from warmer FGK-dwarfs, demonstrating that the APOGEE spectra can be used to Galactic chemical evolution using large samples of selected M-dwarfs.
\end{abstract}


\section{Introduction}

M dwarfs are low-mass stars (the lowest-mass M dwarfs falling just above the hydrogen-burning limit), with small radii, and comprising roughly 70\% of Galactic stellar populations (\citealt{Miller1979}, \citealt{Henry2018}). 
Because of their relatively small masses and sizes, M dwarfs are intrinsically faint stars that are challenging to study at large distances using telescopes currently available;  nonetheless, due to their large numbers, a great many M dwarfs in the solar neighborhood have been studied in the optical and near-infrared (NIR) at both low and high spectral resolutions.

The optical spectra of M dwarfs contain several strong molecular absorption features (e.g., TiO, or VO; \citealt{Allard2000}) which limit the use of 'well-known' spectral analysis techniques that are used typically in high-resolution abundance studies of the warmer FKG main-sequence stars. Despite this difficulty, a number of works using optical lines that are likely free of molecular blends (e.g., \citealt{WoolfWallerstein2005a}, \citealt{Bean2006}, and  \citealt{WoolfWallerstein2020}) have shown that it is possible to derive individual abundances of Fe, Ca, and Ti via spectrum synthesis analyses of high-resolution optical M dwarf spectra.
Also using high-resolution optical spectra, \cite{ChavezLambert2009} derived $^{46}$Ti/$^{50}$Ti isotopic ratios in eleven M dwarfs in the halo and the disk (thin and thick), as well as analyzing TiO lines to confirm their metallicity scale.
\cite{Veyette2017} used equivalent widths of Fe I and Ti I lines from $Y$-band spectra to produce a calibration for $T_{\rm eff}$, [Fe/H], and [Ti/Fe], achieving an internal precision of 60 K, 0.10, and 0.05 dex, respectively, in their results, which are typical of those obtained for the warmer F, G, and K stars. 

Studying M dwarfs in the NIR brings two main advantages to the analysis: they are much brighter in the NIR than in the optical, and, in general, the NIR spectra of M dwarfs have fewer strong molecular blends than in their optical spectra.  Although molecular lines of CO, H$_{2}$O, and FeH are present in the NIR, their absorption is not as dense and deep; note, however, that  millions of water lines in the APOGEE (Apache Point Observatory Galactic Evolution Experiment \citealt{Majewski2017}) $H$-band become dominant for M dwarfs with low effective temperatures ($T_{\rm eff}$ $\leq$ 3500 K ).
In the previous works of \cite{TsujiNakagima2014}, \cite{Tsuji2015}, \cite{TsujiNakagima2016}, C and O abundances were determined via spectral syntheses in the $K$-band using the molecular lines of CO and H$_{2}$O. Also using the $K$-band, \cite{RojasAyala2012} developed a spectroscopic calibration based on equivalent widths of Na I and Ca I lines to determine M dwarf metallicities (see also \citealt{Covey2010}, \citealt{RojasAyala2010}, \citealt{Mann2013b}, \citealt{Newton2014}, \citealt{Terrien2015}). 

From high-resolution $J$-band spectra, \cite{Lindgren2016} and \cite{Lindgren2017} demonstrated that FeH and Fe I lines can be used to determine precise exoplanet hosting M dwarf metallicities via spectrum synthesis.
The CARMENES and HPF surveys (\citealt{Carmenes2014}, \citealt{Mahadevan2012}) obtains high-resolution spectra covering both NIR and optical spectral regions: this spectral combination constitutes one of the most powerful tools to characterize M dwarf spectra (see \citealt{Passegger2018} and \citealt{Reiners2018}). From an analysis of CARMENES spectra in the region between roughy $\lambda$9,000--13,000\AA, \cite{Ishikawa2020} determined abundances of eight elements for a sample of five M dwarfs in binary systems. \cite{Ishikawa2021} used the Subaru/IRD high-resolution spectra to determine detailed chemical abundances for 13 M dwarfs.

The high-resolution spectroscopic APOGEE survey has also opened a new window into the chemical analysis and stellar characterization of significant numbers of M dwarfs in the Milky Way. Using different approaches,  \cite{Rajpurohit2018}, \cite{Birky2020}, \cite{Souto2020}, and \cite{Sarmento2021} determined the atmospheric parameters and metallicities of APOGEE M dwarfs, while previous work by \cite{Souto2017,Souto2018Ross} demonstrated proof-of-concept analysis techniques that yielded individual abundances for more than ten chemical elements from the $H$-band APOGEE spectra.

In this work we present the detailed chemical abundance study for a sample of 21 M dwarf stars. We determined the chemical abundances of fourteen elements covering different nucleosynthetic origins: C (produced primarily in low- to intermediate-mass stars); the $\alpha$-elements O, Mg, Si, and Ca; the odd-Z elements Na, Al, and K; and the iron-peak elements Cr, Mn, Fe, and Ni; all derived using the APOGEE spectra and an updated APOGEE line list \citep{Smith2021}. The targets were taken from our previous studies \citep{Souto2020,Souto2021}, as they could serve as benchmark stars for metallicity and effective temperature calibrations and validations. Here, with the addition of detailed chemistry for these targets, which are for the most part solar neighborhood M dwarf members of binary systems, or benchmarks with measured angular diameters, 
we can begin to compare chemical abundances for M dwarfs with other populations in the Milky Way and investigate their behavior in the canonical [X/Fe] versus [Fe/H] plane. 
The M dwarf abundances can also be used to calibrate results in DR16, which were derived automatically using the APOGEE ASPCAP pipeline \citep{Jonsson2020}. 
Section 2 of this paper presents the studied sample, Section 3 the abundance analysis, and the determination of the stellar parameters. In Section 4, we discuss the results and summarize them in Section 5.

\section{The Data and the Sample}

The M dwarfs analyzed here were taken from \cite{Souto2020} and \cite{Souto2021}. This sample contains eleven M dwarf stars in wide binary systems with warmer primaries selected from \cite{Mann2013_binarypaper} and \cite{Montes2018}, along with ten field M dwarfs with interferometric radii measured by \cite{Boyajian2012}.

The targets were observed as part of the APOGEE-1 / SDSS-IV (Sloan Digital Sky Survey; \citealt{SDSS4}) from observations using the APOGEE-N spectrograph, a multi-fiber (300) high-resolution (R=$\lambda$/$\Delta$$\lambda$$\sim$22,500; \citealt{Gunn2006}, \citealt{Wilson2019}) cryogenic spectrograph operating in the near-infrared $H$-band ($\lambda$1.51 -- 1.69$\micron$) and located at the Apache Point Observatory (New Mexico, USA). 
Here, we use updated reduced spectra for the sample \citep{Nidever2015} from the publicly available 16$th$ SDSS data release (DR16, \citealt{DR16}, \citealt{Jonsson2020}).

\begin{figure*}
\begin{center}
  \includegraphics[angle=90,width=0.95\linewidth,clip]{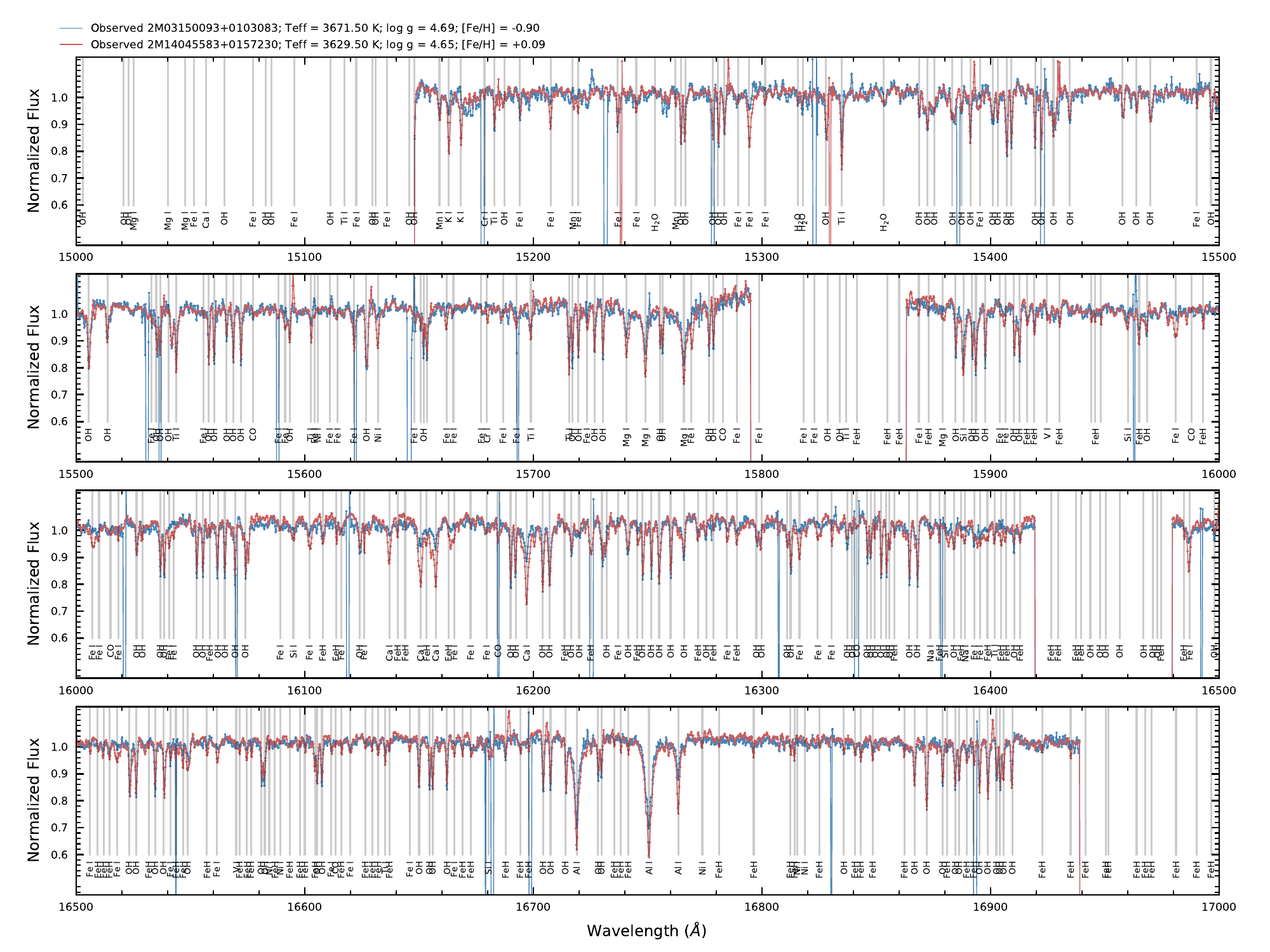}
\caption{Line identification in the normalized APOGEE spectra of a solar metallicity (2M14045583+0157230) and a metal-poor (2M03150093+0103083) M dwarf represented as red and blue lines, respectively. All spectral lines available in the M dwarf APOGEE spectra are shown as a gray line.}
\end{center}
\label{spectrum_patronus}
\end{figure*}

\begin{figure*}
\begin{center}
\includegraphics[angle=0,width=1.0\linewidth,clip]{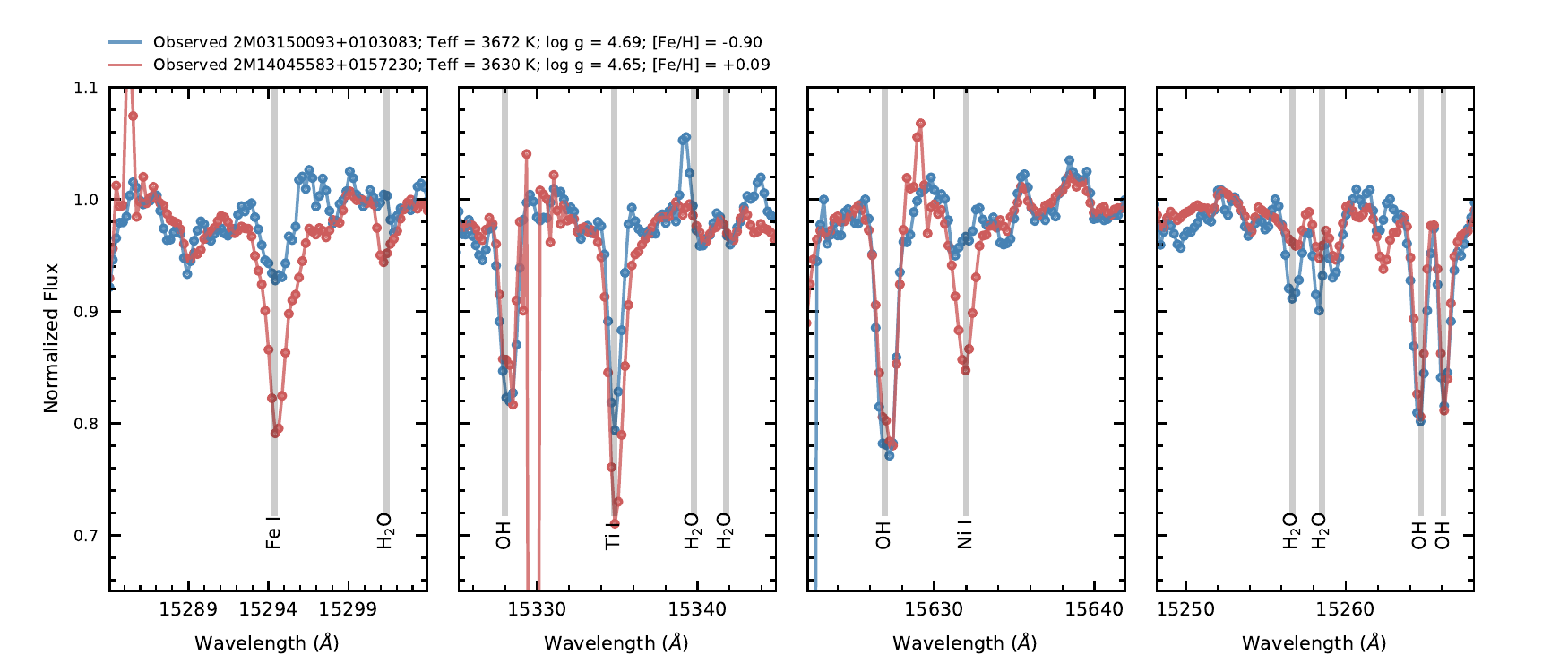}
\caption{Expanded views of four lines in Figure \ref{spectrum_patronus}. The panels display the $\lambda$15294.4 \AA{} Fe I, $\lambda$15334.8 \AA{} Ti I, $\lambda$15632.0 \AA{} Ni I, and $\lambda$15258.1 \AA{} H$_{2}$O, from left to right, respectively.}
\end{center}
\label{spectrum_region}
\end{figure*}

\section{Stellar Parameters and Abundance Analysis}

The effective temperatures ($T_{\rm eff}$), surface gravities (log $g$), metallicities ([Fe/H]), along with carbon (A(C)) and oxygen (A(O)) abundances for the studied stars, were determined using solutions to $T_{\rm eff}$--A(O) pairs and log $g$--A(O) pairs; the method consists in deriving oxygen abundances from different abundance indicators (OH and H$_{2}$O lines) as functions of $T_{\rm eff}$ and log $g$. This procedure leads to a unique value of $T_{\rm eff}$ and log $g$ yielding the same oxygen abundance derived from the different oxygen abundance indicators. We refer the reader to \cite{Souto2020} for more details on the methodology.

The abundance calculations are based on the 1D plane-parallel LTE MARCS model atmospheres (\citealt{Gustafsson2008}), the turbospectrum code (\citealt{AlvarezPLez1998} and \citealt{Plez2012}) and the BACCHUS wrapper (\citealt{Masseron2016}) in the manual mode. 
Synthetic spectra were generated using the most recent version of the APOGEE line list \citep{Smith2021}, which was also used in DR17. 
We determined chemical abundances via comparisons of synthetic and observed spectra. 

Individual abundances were measured, whenever possible, for a total 113 spectral lines: 15 lines of Fe I, 25 of FeH, 2 of CO, 4 of H$_{2}$O, 32 of OH, 2 of Na I, 3 of Mg I, 3 of Al I, 4 of Si I, 2 of K I, 3 of Ca I, 6 of Ti I, 1 of V I, 1 of Cr I, 3 of Mn I based on \cite{Souto2018Ross} (stars with $T_{\rm eff}$ $<$ 3500 K) and \cite{Souto2017} (stars with $T_{\rm eff}$ $>$ 3500 K).  
In this work, we add seven Ni I lines at air wavelengths of $\lambda$15605.7\AA, $\lambda$15632.7\AA, $\lambda$16584.4\AA, $\lambda$16589.3\AA, $\lambda$16673.7\AA, $\lambda$16815.5\AA, $\lambda$16818.8\AA{}, which are measurable in the APOGEE spectra of metal-rich M dwarfs but not previously studied in \cite{Souto2017} and \cite{Souto2018Ross}, as these lines are very weak for solar and sub-solar metallicities.
The Ti and V abundances derived here displayed an unexpected trend as a function of metallicity. Therefore, we remove these species from the discussion section as work needs to be done to determine what is driving this offset.
The adopted atmospheric parameters and the derived individual abundances are presented in Tables 1 and 2, respectively.

\begin{deluxetable*}{llrrrrrrr}
\tablenum{1}
\tabletypesize{\scriptsize}
\tablecaption{Stellar Parameters}
\tablewidth{0pt}
\startlongtable
\tablehead{
\colhead{2Mass ID} &
\colhead{ID} &
\colhead{J} &
\colhead{H} &
\colhead{Ks} &
\colhead{d(pc)} &
\colhead{$T_{\rm eff}$} &
\colhead{log $g$} &
\colhead{[Fe/H]}}
\startdata
Binaries \\
2M03044335+6144097  &GJ 3195	&8.877	&8.328	&8.103	&23.5 	 &3541	&4.72&	-0.33	 \\
2M03150093+0103083	&NLTT 10349  &11.622	&11.043	&10.855	&77.5 	 &3672	&4.69&   -0.92 \\
2M03553688+5214291	&LSPM J0355+5214 &10.885	&10.325	&10.127	&39.6 	 &3455	&4.89&   -0.34 \\
2M06312373+0036445	&NLTT 16628  &11.077	&10.465	&10.252	&72.1 	 &3752	&4.70&   -0.39 \\
2M08103429-1348514	&GJ 9255 B   &8.276	&7.672	&7.418	&20.9 	 &3566	&4.76&    0.01\\
2M12045611+1728119	&LSPM J1204+1728S    &9.793	&9.183	&8.967	&37.6 	 &3369	&4.82&   -0.45 \\
2M14045583+0157230	&NLTT 36190  &10.129	&9.483	&9.269	&51.8 	 &3630	&4.65&    0.05\\
2M18244689-0620311	&PM J18247-0620  &9.659	&9.052	&8.795	&39.6 	 &3425	&4.77&    0.05\\
2M20032651+2952000	&GJ 777 B    &9.554	&9.026	&8.712	&16.0 	 &3295	&5.05&    0.21\\
2M02361535+0652191	&GJ 105 B    &7.333	&6.793	&6.574	&7.2  	 &3335	&4.95&   -0.20\\
2M05413073+5329239	&GJ 212  &6.586	&5.963	&5.759	&12.3 	 &3783	&4.74&    0.21 \\
\hline                                                
Interferometric radii \\                              
2M11032023+3558117	&GJ 411  &4.203	&3.640	&3.254	&2.6  	 &3579	&4.81&  -0.44  \\
2M11052903+4331357	&GJ 412 A    &5.538	&5.002	&4.769	&4.8  	 &3567	&4.84&  -0.49  \\
2M00182256+4401222	&GJ 15 A &5.252	&4.476	&4.018	&3.6  	 &3585	&4.85&  -0.39  \\
2M05312734-0340356	&GJ 205  &4.999	&4.149	&4.039	&5.7  	 &3820	&4.67&   0.29 \\
2M09142298+5241125	&GJ 338 A    &4.889	&3.987	&3.988	&6.3  	 &3931	&4.66&   0.02 \\
2M09142485+5241118	&GJ 338 B    &4.779	&4.043	&4.136	&6.3  	 &3870	&4.70&   0.11 \\
2M13454354+1453317	&GJ 526  &5.181	&4.775	&4.415	&5.4  	 &3714	&4.71&  -0.36  \\
2M18424666+5937499	&GJ 725 A    &5.189	&4.741	&4.432	&3.5  	 &3505	&4.84&  -0.34  \\
2M18424688+5937374	&GJ 725 B    &5.721	&5.197	&5.000	&3.5  	 &3400	&4.90&  -0.36  \\
2M22563497+1633130	&GJ 880  &5.36	&4.800	&4.253	&6.9    &3643	&4.78&   0.26 \\
\enddata
\tablenotetext{}{The estimated uncertainties in $T_{\rm eff}$, log $g$, and [Fe/H] are about 100K, 0.20 dex, and 0.10 dex, respectively. }
\end{deluxetable*}

Figure \ref{spectrum_patronus} shows the APOGEE spectra (from $\sim\lambda$1.5 to 1.7$\micron$) for two M dwarfs in our sample. The two selected stars have similar stellar parameters ($T_{\rm eff}$ and log $g$) but different metallicities;
while 2M14045583+0157230 is solar metallicity (red line), 2M03150093+0103083 is metal-poor ([Fe/H] $\sim$ -1.0; blue line).
It is apparent from a comparison of their spectra that the overall shape of both spectra is similar, which is driven by the similarity of their values of $T_{\rm eff}$ and log $g$.
However, the strengths of the absorption lines are shallower/weaker for several species in the metal-poor star. This behaviour is more prominent for the neutral atomic lines.

Figure \ref{spectrum_region} presents an expansion of the region in Figure \ref{spectrum_patronus} containing four lines: From left to right $\lambda$15294.4\AA{} Fe I, the $\lambda$15334.8\AA{} Ti I, the $\lambda$15632.0\AA{} Ni I, and $\lambda$15258.1\AA{} H$_{2}$O.
In the metal-poor M dwarf, the neutral lines have absorption depth with roughly 10--15\% lower depths when compared to the solar-metallicity M dwarf, while the OH lines display roughly the same depths for both stars, and  the  H$_{2}$O lines in the metal-poor star are stronger than those solar metallicity one, which is expected if the metal-poor star has an enhanced value of [O/Fe].

\subsection{Abundance Uncertainties}

The uncertainties in the derived abundances in this work are estimated to be the same as in our previous investigations, given the similarities in the adopted analysis methodologies. For the early M dwarfs ($T_{\rm eff}$ $>$ 3500 K) we refer to the uncertainties presented in \cite{Souto2017}, and for the mid-type M dwarfs ($T_{\rm eff}$ $<$ 3500 K see \cite{Souto2018Ross}. 
The uncertainties in the derived atmospheric parameters are the same as the ones presented in \cite{Souto2020}: $\pm$ 100K and $\pm$ 0.16 dex for $T_{\rm eff}$ and log $g$, respectively.

\section{Results and Discussion}

Stellar parameters and iron abundances for the stars in this study were previously derived in \cite{Souto2020}. However, this previous study used spectra from a previous version of the APOGEE data reduction pipeline \citep{Jonsson2020}, along with a previous version of the line list \citep{Shetrone2015}, while here we adopt the most recent APOGEE line list that has several updates \citep{Smith2021}. 
A comparison of the stellar parameters and metallicities obtained with those derived previously from the same APOGEE spectra finds good agreement: delta $T_{\rm eff}$ = 16 $\pm$ 56 K; delta log $g$ = 0.00 $\pm$ 0.10 dex, and delta [Fe/H] = 0.10 $\pm$ 0.04 dex. 
Assuming only the interferometric stars we obtain: delta $T_{\rm eff}$ = 8 $\pm$ 79 K; delta log $g$ = 0.03 $\pm$ 0.10 dex.
In this study the derived metallicities are the mean of the iron abundances from the Fe I and FeH lines, while \cite{Souto2020} used only Fe I lines as their metallicity indicator (see discussion in \citealt{Souto2021}).

Three stars of this work have exoplanets detected: GJ 411, GJ 15 A, and GJ 338 B.
GJ 411 hosts a ``mass-like'' Earth with M$_{\Earth}$*sin($i$) = 2.69, equilibrium temperature ($T_{\rm eq}$) and insolation flux ($S_{\Earth}$) of 370 K, and 3.13, respectively (\citealt{Stock2020}). GJ 15 A has two exoplanets detected to date (\citealt{Pinamonti2018}), where GJ 15 A b is considerably less massive than GJ 15 A c, M$_{\Earth}$*sin($i$) = 3.03, and 36. 
GJ 338 B likely hosts a super Earth, where GJ 338 B b has M$_{\Earth}$*sin($i$) = 10.27 and $T_{\rm eq}$ = 391 K (\citealt{GonzalezAlvarez2020}).
These four exoplanets were detected by radial velocity, and no transit observations are available to date.
This work abundance results can be used to study the star-planet connection and geophysical properties of these likely rocky exoplanets in future works.

\begin{deluxetable*}{lrrrrrrrrrrrrrrrrrrrrrr}
\tablenum{2}
\tabletypesize{\tiny}
\tablecaption{Stellar Abundances}
\tablewidth{0pt}
\startlongtable
\tablehead{
\colhead{Element} &
\colhead{4097} &
\colhead{3083} &
\colhead{4291} &
\colhead{6445} &
\colhead{8514} &
\colhead{8119} &
\colhead{7230} &
\colhead{0311} &
\colhead{2000} &
\colhead{2191} &
\colhead{9239} &
\colhead{8117} &
\colhead{1357} &
\colhead{1222} &
\colhead{0356} &
\colhead{1125} &
\colhead{1118} &
\colhead{3317} &
\colhead{7499} &
\colhead{7374} &
\colhead{3130} 
}
\startdata
$\langle$A(Fe[Fe I])$\rangle$	&7.20	&6.55	&7.16	&7.09	&7.50	&7.05	&7.54	&7.50	&7.70	&7.32	&7.69	&7.05	&7.05	&7.10	&7.81	&7.52	&7.59	&7.14	&7.10	&7.13	&7.81	&\\
$\langle$A(Fe[FeH])$\rangle$	&7.04	&6.51	&7.06	&7.04	&7.43	&6.95	&7.47	&7.50	&7.62	&7.18	&7.63	&6.97	&6.87	&7.03	&7.67	&7.42	&7.52	&7.04	&7.12	&7.05	&7.61	&\\
$\langle$A(C)$\rangle$	&8.20	&7.62	&8.03	&8.10	&8.34	&8.01	&8.40	&8.36	&8.62	&8.29	&8.55	&8.04	&7.92	&7.96	&8.65	&8.26	&8.32	&8.24	&8.18	&8.19	&8.58	&\\
$\langle$A(O[H2O])$\rangle$	&8.58	&8.21	&8.39	&8.47	&8.62	&8.31	&8.65	&8.66	&8.86	&8.61	&8.80	&8.44	&8.30	&8.33	&8.89	&8.52	&8.60	&8.59	&8.54	&8.52	&8.80	&\\
$\langle$A(O[OH])$\rangle$	&8.60	&8.20	&8.40	&8.48	&8.62	&8.30	&8.65	&8.68	&8.86	&8.63	&8.76	&8.44	&8.32	&8.34	&8.85	&8.56	&8.58	&8.57	&8.53	&8.55	&8.80	&\\
$\langle$A(Na)$\rangle$	&5.95	&...&...&5.80	&6.17	&...&6.25	&6.19	&...&...&6.27	&...&...&...&6.71	&6.14	&...&...&...&...&6.55	&\\
$\langle$A(Mg)$\rangle$	&7.62	&7.11	&7.33	&7.29	&7.62	&7.27	&7.58	&7.72	&7.55	&7.56	&7.60	&7.37	&7.35	&7.38	&7.74	&7.41	&7.56	&7.39	&7.29	&7.51	&7.95	&\\
$\langle$A(Al)$\rangle$	&6.22	&5.62	&5.95	&5.95	&6.24	&5.84	&6.32	&6.39	&6.50	&6.19	&6.37	&6.01	&5.96	&6.02	&6.60	&6.17	&6.30	&6.11	&6.07	&6.24	&6.77	&\\
$\langle$A(Si)$\rangle$	&7.37	&6.83	&7.24	&7.25	&7.47	&...&7.52	&...&...&...&7.61	&7.22	&7.17	&7.35	&7.75	&7.45	&7.58	&7.28	&7.16	&...&7.89	&\\
$\langle$A(K)$\rangle$	&4.97	&4.42	&4.82	&4.80	&4.99	&4.64	&5.04	&5.04	&5.22	&4.93	&5.07	&4.68	&4.67	&4.71	&5.31	&4.93	&4.95	&4.88	&4.85	&4.77	&5.25	&\\
$\langle$A(Ca)$\rangle$	&6.29	&5.76	&6.06	&6.14	&6.39	&6.02	&6.33	&6.40	&6.39	&6.32	&6.45	&6.08	&6.07	&6.16	&6.59	&6.27	&6.28	&6.13	&6.10	&6.15	&6.60	&\\
$\langle$A(Cr)$\rangle$	&...&...&...&5.33	&5.65	&...&5.70	&...&...&...&5.87	&...&...&...&6.08	&5.66	&5.72	&...&...&...&6.04	&\\
$\langle$A(Mn)$\rangle$	&5.09	&...&...&4.83	&5.40	&...&5.36	&...&...&...&5.49	&4.78	&4.85	&4.90	&5.70	&5.25	&5.33	&4.98	&...&...&5.68	&\\
$\langle$A(Ni)$\rangle$	&...&...&...&...&...&...&...&...&...&...&6.42	&...&...&...&6.68	&6.28	&6.39	&...&...&...&...	&\\
\tablewidth{0pt}	
\enddata
\end{deluxetable*}

\subsection{Verifying the Abundance Scale using M Dwarfs in Binary Systems}

As discussed previously, M dwarfs that are members of wide binary systems provide an opportunity to compare their abundances with those for the warmer primaries, which can be obtained from a number of studies in the literature (\citealt{Ishikawa2020}, \citealt{Montes2018}, \citealt{Mann2013_binarypaper}, \citealt{Bonfils2005}). 
There remains one caveat in the use of binary systems as abundance tests for M dwarfs: the assumption that the chemical compositions of the M dwarf and its primary are identical.  The process of atomic diffusion will alter the surface abundances, to varying degrees, of main sequence stars over time \citep{Michaud2015}, with heavy elements diffusing downward, out of the outer convective envelope, relative to H, thus lowering values of [X/H] in the photosphere relative to their initial values on the ZAMS. The magnitude of the abundance changes wrought by diffusion are strong functions of stellar mass and age, with changes being of the order of 0.01 to $\sim$0.1 dex.  Atomic diffusion has been observed in the abundance distributions in the open clusters M67 (\citealt{Souto2018,Souto2019}, \citealt{Gao2018}, \citealt{BertelliMotta2018}), NGC 2420 \citep{Semenova2020}, and Coma Berenices \citep{Souto2021}.  As M dwarfs have relatively massive and deep convective envelopes, when compared to hotter (and more massive) main sequence stars, diffusion effects are much reduced.  In a binary system consisting of an M dwarf and a hotter and more massive FGK primary, diffusion will act preferentially over time to reduce the heavy-element abundances in the primary relative to the M dwarf companion.  

The warm primaries in this study cover a range in effective temperature from roughly 4600 K to 6300 K and in principle, they would experience varying levels of diffusion (0.01 to 0.12 dex assuming solar age; \citealt{Choi2016}). However, even if the amount of diffusion that the primaries suffer is unknown, diffusion effects are rough similar for the different elements studied (\citealt{Pinsonneault1989}, \citealt{Choi2016}, \citealt{Dotter2017}) and this similar behavior can be used in the form of [X/Fe] ratios, which, to a degree, removes and  minimizes diffusion effects. 
We emphasize, however, that we do not aim to measure diffusion effects in binary systems. We are simply analyzing the El/Fe ratio, which would tend to erase diffusion signatures (if they exist), as all elements are affected by roughly the same level of photospheric abundance depletion (\citealt{Choi2016}).

\begin{figure*}
\begin{center}
\includegraphics[angle=0,width=0.95\linewidth,clip]{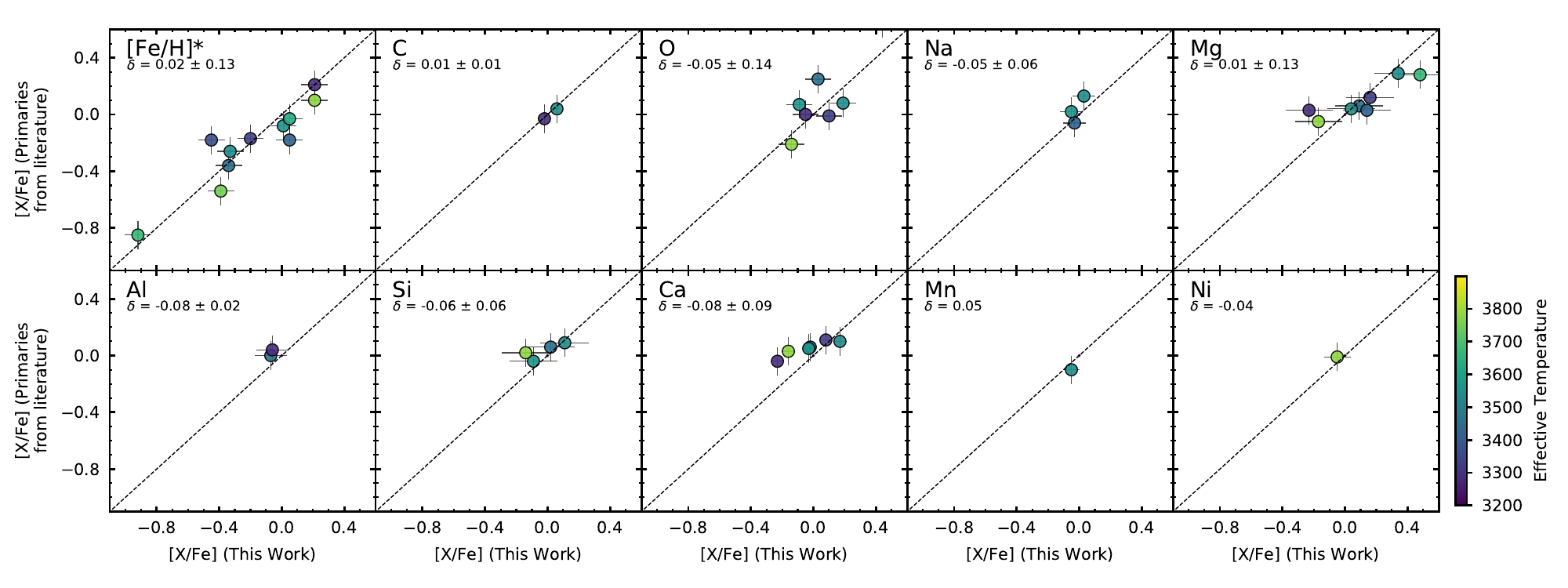}
\includegraphics[angle=0,width=0.95\linewidth,clip]{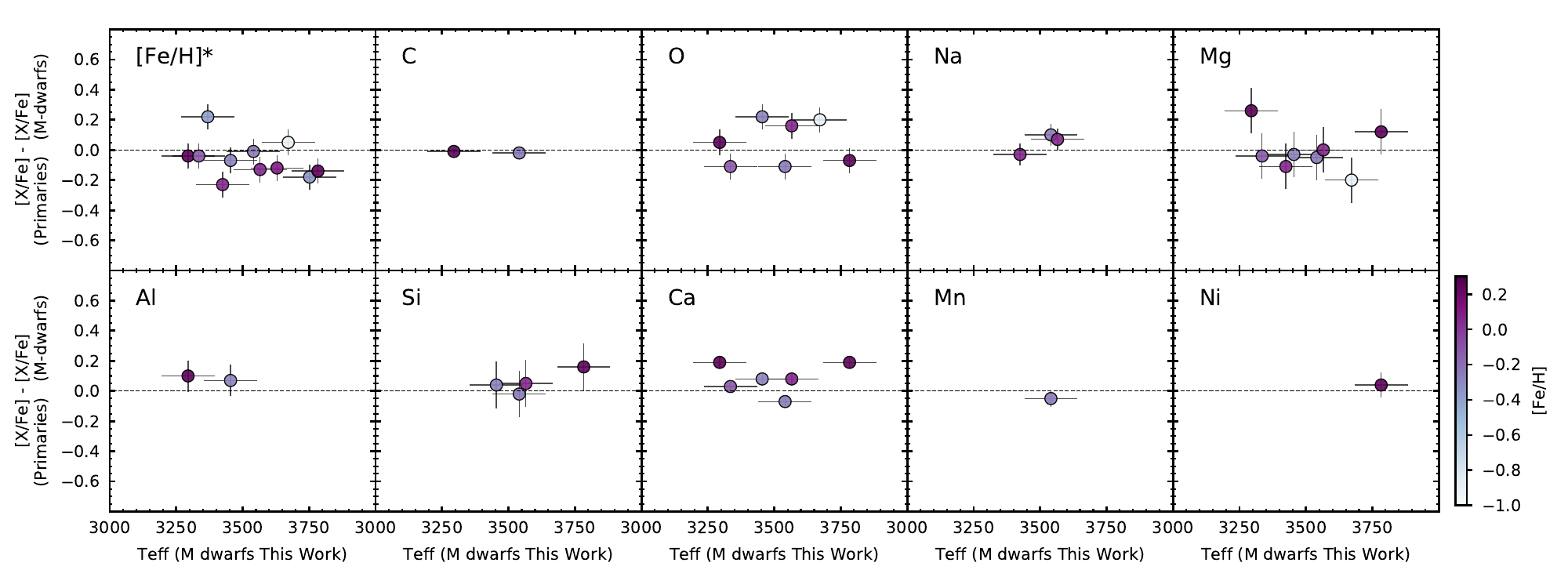}
\caption{Top panels: comparison of the [X/Fe] abundances in binary systems containing an M dwarf and a warmer primary of type FGK;. Bottom panels: the residual [X/Fe] (primaries) -- [X/Fe] (M dwarfs) versus the M dwarf effective temperatures. For iron we adopt [Fe/H] instead of [X/Fe].}
\end{center}
\label{XH_XH}
\end{figure*}

Comparisons of our M dwarf [X/Fe] results with literature values of [X/Fe] for the primaries for ten elements, are shown in the top panels of Figure \ref{XH_XH}, and the offsets are summarized in Table 3. The mean differences between the elemental abundances-relative-to-iron of the primary stars minus M dwarfs in this work (along with standard deviations) are labeled in the top left of each panel (see Table 3).
The bottom panels of the Figure display the corresponding residual diagram ([X/Fe] (warm primaries) -- [X/Fe] (M dwarfs) plotted as a function of our derived effective temperature. 
Overall, these abundances for the M dwarfs are in good/reasonable agreement with the comparison values from the literature (see Table 3).
The comparison of individual elemental abundances between the M dwarfs and their companion primary stars is limited for many of the elements by published elemental abundances for the primaries. The elements K and Cr, for example, do not have literature results for the primary stars.

The iron abundances are presented relative to H (top left panel of Figure 3), and therefore could be affected to some level by diffusion (see \citealt{Dotter2017}). Possible effects of diffusion aside, there is almost no offset between the mean abundances $\langle$[Fe/H] (this work) - [Fe/H] (warm primaries)$\rangle$ = +0.02 $\pm$ 0.13 dex. (\citealt{Souto2020} used an older APOGEE line list and found a slightly different mean metallicity difference by 0.02 dex.)
Iron is the best represented abundance in the comparisons, as all eleven binary systems have published values of [Fe/H]. The standard deviation of the mean for [Fe/H] of $\pm$0.13 dex is a typical level of dispersion found in abundance analyses, given that all abundances carry some level of systematic uncertainties and internal errors, and, because they come from different methodologies and literature sources (see \citealt{Hypatia}).

The elements O, Mg, Si, and Ca are also represented by more than just a few comparison binary systems (N$\ge$4). Using values of [X/Fe] as references for the abundance comparisons, the differences are quite small and the standard deviations are 0.14 dex and below, as can be seen in Figure \ref{XH_XH}. The mean abundance differences in "Primaries - M dwarfs" is -0.05 $\pm$ 0.03 dex.  Taken together, the results for the O, Mg, Si, and Ca indicate that there are no significant trends between the M dwarf abundance scale and the warmer FGK companion stars.

The remaining elements are represented in the binary sample by only small numbers of comparisons, so any underlying trends in the M dwarf abundances would be difficult to discern, unless they are systematic and large. 
For example, only two sample stars had carbon abundance results in the literature for their primaries. The mean abundance difference (M dwarf -- Primary) is $\delta$ = 0.01 $\pm$ 0.01 dex. Aluminum and sodium also had only two or three comparison systems each, with no striking conclusions to be drawn, other than that there are no large differences found.  

The final two elements to be discussed are Mn and Ni, where only one literature result for the primary stars is found  to compare with the M dwarf abundance.  These elements are included largely for completeness; the differences for these single comparisons of $\delta$ [Mn/Fe] = 0.05, and $\delta$ [Ni/Fe] = -0.04 dex.

\begin{deluxetable*}{lrrrrrrrrrrrr}
\rotate
\tablenum{3}
\tabletypesize{\scriptsize}
\tablecaption{Abundance ratios}
\tablewidth{0pt}
\startlongtable
\tablehead{
\colhead{M dwarfs secondary} &
\colhead{4097} &
\colhead{3083} &
\colhead{4291} &
\colhead{6445} &
\colhead{8514} &
\colhead{8119} &
\colhead{7230} &
\colhead{0311} &
\colhead{2000} &
\colhead{2191} &
\colhead{9239}\\
\colhead{Warmer primaries} &
\colhead{HIP 14286} &
\colhead{HIP 15126} &
\colhead{HIP 18366} &
\colhead{HIP 31127} &
\colhead{HIP 40035} &
\colhead{HIP 58919} &
\colhead{HIP 68799} &
\colhead{HIP 90246} &
\colhead{HIP 98767} &
\colhead{HIP 12114} &
\colhead{HIP 26779}
}
\startdata
$\langle$[Fe/H]$\rangle$ (M dwarfs)	&-0.33	&-0.92	&-0.34	&-0.39	&0.01	&-0.45	&0.05	&0.05	&0.21	&-0.20	&0.21\\
$\langle$[Fe/H]$\rangle$ (Warmer primaries)	&-0.26	&-0.85	&-0.36	&-0.54	&-0.08	&-0.18	&-0.03	&-0.18	&0.21	&-0.17	&0.10\\
$\delta$ (M dwarfs - warmer primaries)	&-0.07	&-0.07	&0.02	&0.15	&0.09	&-0.27	&0.08	&0.23	&0.00	&-0.03	&0.11\\
$\langle$[C/Fe]$\rangle$ (M dwarfs)	&0.14	&0.15	&-0.02	&0.09	&-0.07	&0.07	&-0.04	&-0.08	&0.02	&0.10	&-0.05\\
$\langle$[C/Fe]$\rangle$ (Warmer primaries)	&0.04	&...&...&...&...&...&...&...&-0.03	&...&...\\
$\delta$ (M dwarfs - warmer primaries)	&0.10	&...&...&...&...&...&...&...&0.05	&...&...\\
$\langle$[O/Fe]$\rangle$ (M dwarfs)	&0.27	&0.46	&0.08	&0.21	&-0.06	&0.09	&-0.06	&-0.03	&-0.01	&0.17	&-0.11\\
$\langle$[O/Fe]$\rangle$ (Warmer primaries)	&0.08	&0.64	&0.25	&...&0.07	&...&...&...&0.00	&-0.01	&-0.21\\
$\delta$ (M dwarfs - warmer primaries)	&0.19	&-0.18	&-0.17	&...&-0.13	&...&...&...&-0.01	&0.18	&0.10\\
$\langle$[Na/Fe]$\rangle$ (M dwarfs)	&0.11	&...&...&0.02	&-0.01	&...&0.03	&-0.03	&...&...&-0.11\\
$\langle$[Na/Fe]$\rangle$ (Warmer primaries)	&0.13	&0.02	&-0.04	&...&0.02	&...&...&-0.06	&0.03	&...&...\\
$\delta$ (M dwarfs - warmer primaries)	&-0.02	&...&...&...&-0.03	&...&...&0.03	&...&...&...\\
$\langle$[Mg/Fe]$\rangle$ (M dwarfs)	&0.42	&0.50	&0.14	&0.15	&0.08	&0.19	&0.00	&0.14	&-0.19	&0.23	&-0.14\\
$\langle$[Mg/Fe]$\rangle$ (Warmer primaries)	&0.29	&0.28	&0.06	&...&0.04	&...&...&0.03	&0.03	&0.12	&-0.05\\
$\delta$ (M dwarfs - warmer primaries)	&0.13	&0.22	&0.08	&...&0.04	&...&...&0.11	&-0.22	&0.11	&-0.09\\
$\langle$[Al/Fe]$\rangle$ (M dwarfs)	&0.18	&0.17	&-0.08	&-0.03	&-0.15	&-0.08	&-0.11	&-0.03	&-0.08	&0.02	&-0.21\\
$\langle$[Al/Fe]$\rangle$ (Warmer primaries)	&...&...&0.00	&...&...&...&...&...&0.04	&...&...\\
$\delta$ (M dwarfs - warmer primaries)	&...&...&-0.08	&...&...&...&...&...&-0.12	&...&...\\
$\langle$[Si/Fe]$\rangle$ (M dwarfs)	&0.19	&0.24	&0.07	&0.13	&-0.05	&...&-0.04	&...&...&...&-0.11\\
$\langle$[Si/Fe]$\rangle$ (Warmer primaries)	&0.09	&...&0.06	&...&-0.04	&...&...&0.07	&0.09	&...&0.02\\
$\delta$ (M dwarfs - warmer primaries)	&0.10	&...&0.01	&...&-0.01	&...&...&...&...&...&-0.13\\
$\langle$[Ca/Fe]$\rangle$ (M dwarfs)	&0.31	&0.37	&0.09	&0.22	&0.07	&0.16	&-0.03	&0.04	&-0.13	&0.21	&-0.07\\
$\langle$[Ca/Fe]$\rangle$ (Warmer primaries)	&0.10	&...&0.06	&...&0.05	&...&...&...&-0.04	&0.11	&0.03\\
$\delta$ (M dwarfs - warmer primaries)	&0.21	&...&0.03	&...&0.02	&...&...&...&-0.09	&0.10	&-0.10\\
$\langle$[Mn/Fe]$\rangle$ (M dwarfs)	&0.03	&...&...&-0.17	&0.00	&...&-0.08	&...&...&...&-0.11\\
$\langle$[Mn/Fe]$\rangle$ (Warmer primaries)	&-0.10	&...&...&...&...&...&...&...&...&...&...\\
$\delta$ (M dwarfs - warmer primaries)	&0.13	&...&...&...&...&...&...&...&...&...&...\\
$\langle$[Ni/Fe]$\rangle$ (M dwarfs)	&...&...&...&...&...&...&...&...&...&...&-0.02\\
$\langle$[Ni/Fe]$\rangle$ (Warmer primaries)	&0.03	&...&0.00	&...&0.01	&...&...&-0.10	&0.05	&...&-0.01\\
$\delta$ (M dwarfs - warmer primaries)	&...&...&...&...&...&...&...&...&...&...&-0.01\\
\tablewidth{0pt}	
\enddata
\tablenotetext{}{\tablenotetext{}{Adopted abundances from the literature are from \cite{Shi2004}, \cite{Soubiran2005}, \cite{Reddy2006}, \cite{Mishenina2008,Mishenina2013,Mishenina2015}, \cite{DelgadoMena2010}, \cite{Adibekyan2012}, \cite{Ramirez2012,Ramirez2013},  \cite{Carretta2013}, \cite{Mann2013_binarypaper}, \cite{Bensby2014}, \cite{Battistini2015}, \cite{daSilva2015_helio},  and \cite{Suarez-Andres2017}.}}
\end{deluxetable*}

\subsection{Comparison With DR16 ASPCAP Results}

The main goal of the APOGEE survey is the determination of chemical abundances in Galactic red giant stars; this task relies on the APOGEE ASPCAP pipeline \citep{GarciaPerez2016}, which derives automatically stellar parameters and chemical abundances for more than 25 species.  In this context, the ASPCAP pipeline has been optimized to analyze red-giants and not M dwarfs. ASPCAP determines stellar parameters and chemical abundances in two distinct phases. First, seven parameters ($T_{\rm eff}$, log $g$, [M/H], [C/Fe], [N/Fe], [$\alpha$/Fe], and the microturbulent velocity) are derived via a 7D chi-squared minimization over the entire wavelength range of the APOGEE spectra.
In a second step, ASPCAP uses the atmospheric parameters from step one and derives the individual abundances by minimizing differences in selected spectral windows (or pixels) that are sensitive to the respective elemental abundances. 

In the earlier APOGEE data releases (DR10; \citealt{DR10}, DR12; \citealt{DR12}), the atmospheric parameters and chemical abundances of the relatively small number of M dwarfs observed by APOGEE were not reliable due to the lack of model atmospheres and synthetic models for M dwarfs in the ASPCAP libraries (\citealt{Zamora2015}). 
In addition, the line lists for H$_{2}$O (\citealt{Barber2006}) and FeH (\citealt{Hargreaves2010}), which are crucial for modelling M dwarfs, were only introduced in DR14 (\citealt{DR14}) and DR16, respectively.
DR16 had adequate line lists and model atmospheres grids, with $T_{\rm eff}$ ranging from 3000--4000K, log $g$ from +2.5 -- +5.5 dex, and [Fe/H] from -2.5 -- + 1.00 dex, as presented by \cite{Jonsson2020}
However, the abundance windows in ASPCAP remained optimized for red giants and this most recent data release still lacks abundance windows dedicated to M dwarf stars.

\begin{figure*}
\begin{center}
\includegraphics[angle=0,width=0.8\linewidth,clip]{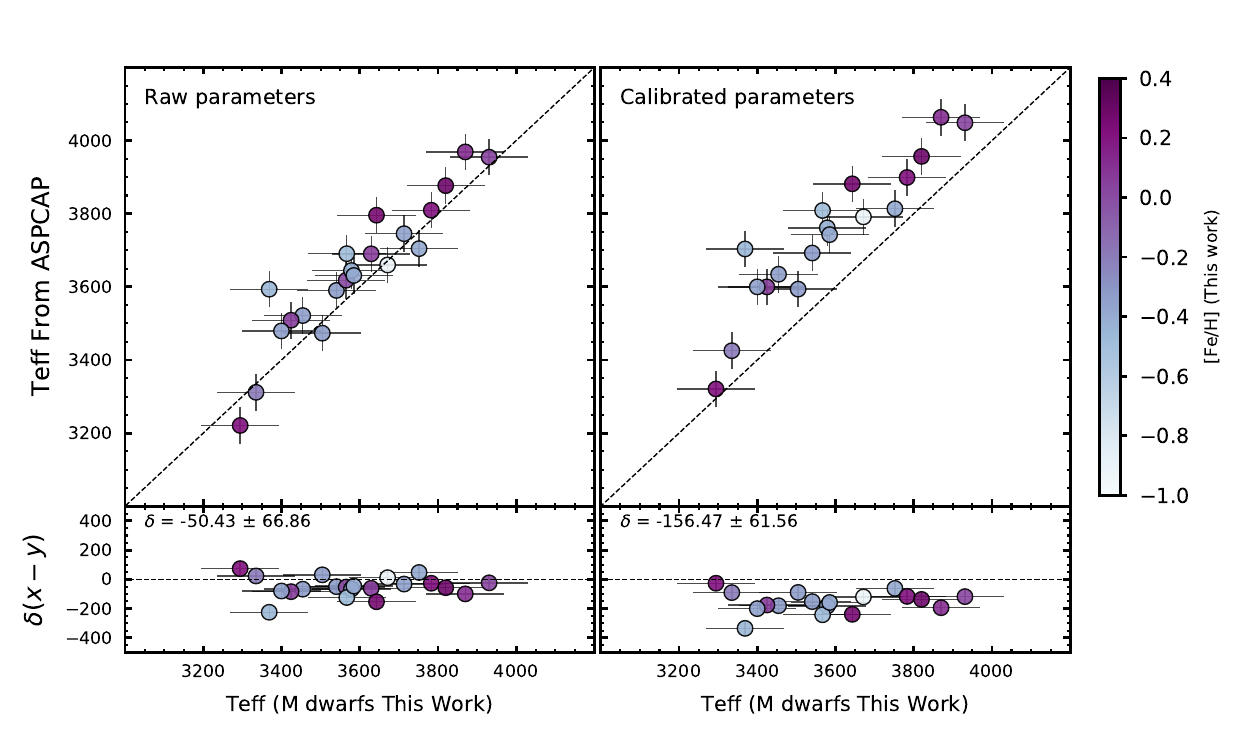}
\includegraphics[angle=0,width=0.8\linewidth,clip]{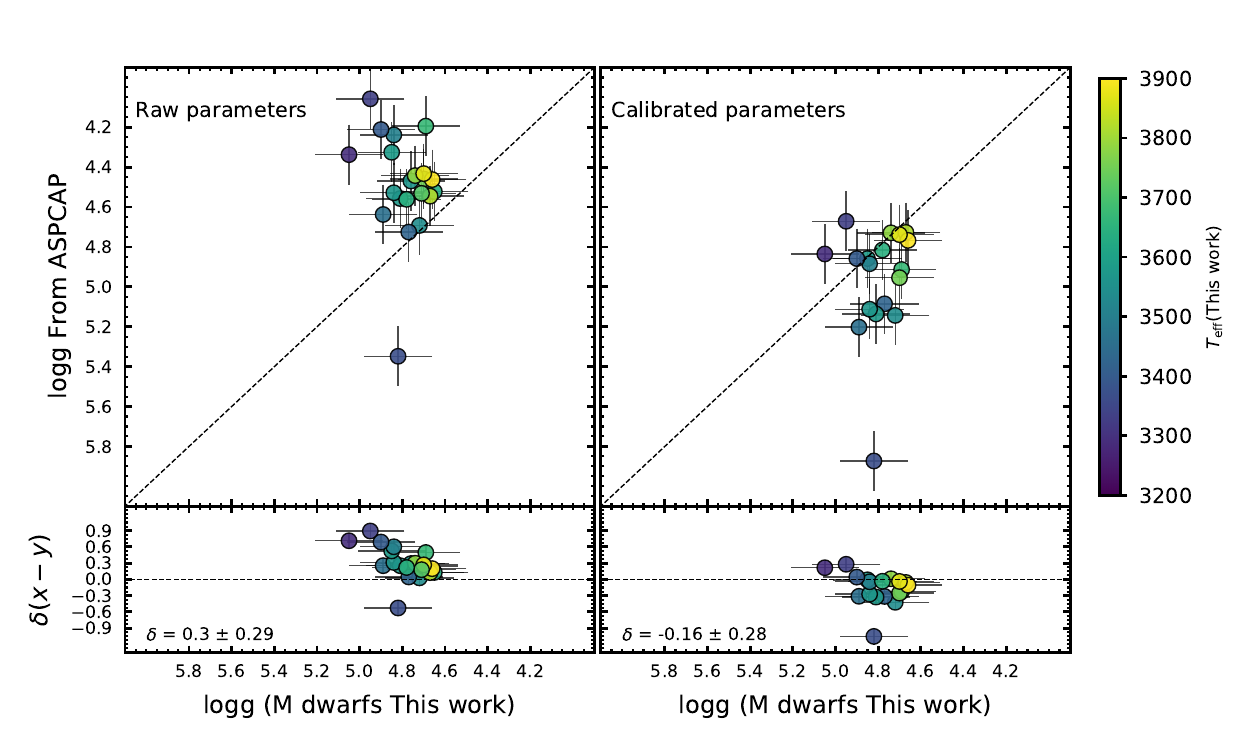}
\caption{The $T_{\rm eff}$--$T_{\rm eff}$ (top panel) and the log $g$--log $g$ (bottom panel) diagrams for this work and DR16 results, respectively. Left and right panels show the raw ASPCAP and the calibrated results from DR16. A residual diagram is presented at the bottom of each panel.}
\end{center}
\label{dr16_atmpar}
\end{figure*}

A comparison between the values of $T_{\rm eff}$ and log $g$ derived here for the M dwarfs with  those from APOGEE DR16 are presented in the top and bottom panels of Figure \ref{dr16_atmpar}, respectively.
This comparison includes both the ASPCAP raw results (left panels) and calibrated ones (right panels, see \citealt{Jonsson2020});  at the bottom of each panel, residual diagrams as functions of $T_{\rm eff}$ and log $g$ are displayed.
There is good agreement between our $T_{\rm eff}$-scale and that of the DR16 ASPCAP raw, where $\langle$$T_{\rm eff, this work}$ - $T_{\rm eff, DR16}$$\rangle$ = -50 $\pm$ 67 K. 
This offset is smaller than the nominal uncertainties in $T_{\rm eff}$, indicating that it is probably not statistically significant. 
The calibrated $T_{\rm eff}$ from the DR16, conversely, show a more significant systematic offset when compared to our results: $\langle$$T_{\rm eff, this work}$ - $T_{\rm eff, DR16}$$\rangle$ = -156 $\pm$ 62 K.

Concerning the surface gravity comparison, we can see from Figure \ref{dr16_atmpar} (left bottom panel) that many of the raw ASPCAP log $g$ values spread around log $g$ $\sim$ 4.8, with some of the values being significantly lower ($\sim$4.4 - 4.1). If we simply compare the mean of the difference we obtain a significant offset and scatter: $\langle$log $g$ $_{\rm this work}$ - log $g$ $_{\rm DR16}$ $\rangle$ = -0.30 $\pm$ 0.29 dex. However, this difference is quite log $g$ dependent, as shown by residual diagram in the bottom-left panel of Figure \ref{dr16_atmpar}; for the coolest M dwarf stars in the sample ($T_{\rm eff}$ $<$ 3400K), the difference $\langle$ log $g$ $_{\rm this work}$ - log $g$ $_{\rm DR16}$ $\rangle$ can reach up to $\sim$ +0.9 dex. 
One star that deviates from this systematic trend is 2M12045611+1728119, for which the surface gravity value derived in DR16 is log $g$ = 5.35 dex and $\delta$ log $g$ = -0.53 dex.
This star is the one in our sample having the highest rotational speed, with v sin $i$ = 13.5 km.s$^{-1}$. 
(The calibrated log $g$ value for this star is log $g$= 5.87 dex, which is even more discrepant; see discussion below.)

It is known that ASPCAP has difficulties in determining good spectroscopic log $g$ values; main-sequence FGK stars and red giant stars present systematic offsets when compared to log $g$'s from physical relations or from seismic results in the literature (e.g., \citealt{Pinsonneault2014,Pinsonneault2018}, \citealt{Holtzman2018}, \citealt{Jonsson2018}, \citealt{Jonsson2020}). Given the biased results for the APOGEE log $g$'s, the APOGEE team calibrates the raw (spectroscopic) ASPCAP log $g$ values and derives calibrated ones (shown in the right bottom panel of Figure \ref{dr16_atmpar}).
For the calibrated log $g$ values, there is an improvement in the comparison with our results; however, the raw ASPCAP log $g$ values appear to have been over-corrected and the spread is still significant when compared to ours; $\langle$ log $g$ $_{\rm this work}$ - log $g$ $_{\rm DR16}$ $\rangle$ = -0.16 $\pm$ 0.28 dex.

\begin{figure*}
\begin{center}
\includegraphics[angle=0,width=0.99\linewidth,clip]{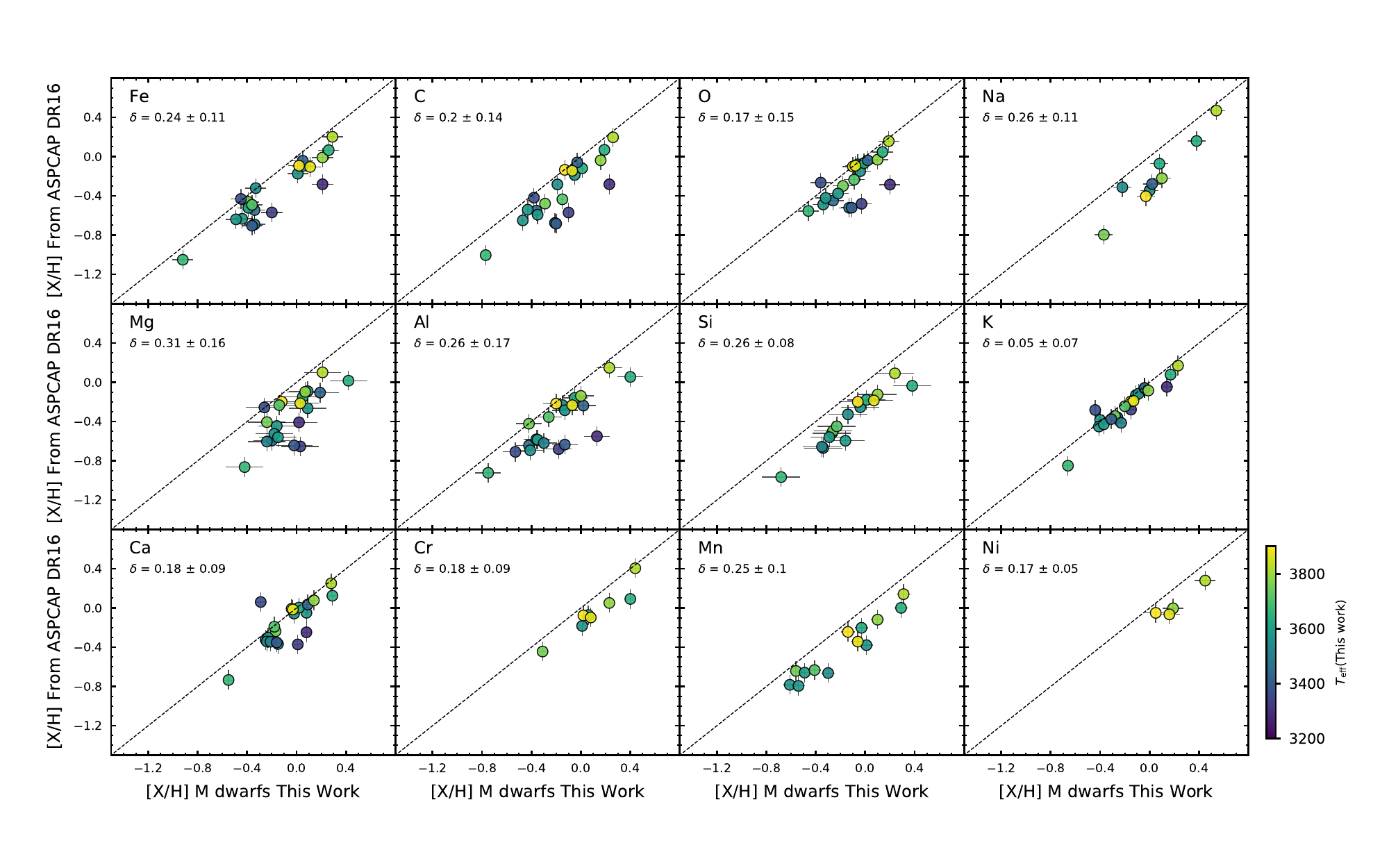}
\caption{The [X/H]--[X/H] diagram with this work M dwarf abundances versus the raw ASPCAP DR16 [X/H]. The delta abundance difference is giving for each element in the panel.}
\end{center}
\label{dr16_abu}
\end{figure*}

The comparison between the elemental abundances from DR16 with this work suggests that, taken as a whole, the ASPCAP abundances are lower than ours $\langle$[M/H] $_{\rm this work}$ - [M/H] $_{\rm DR16}$ $\rangle$ = +0.21 $\pm$ 0.07 dex. 
The comparison for each studied element, along with the delta abundance difference for each element are shown in Figure \ref{dr16_abu}.  The mean abundance differences for all elements are around $\sim$ 0.2 dex, with exceptions being the K on one side ($\delta$ = 0.05 $\pm$ 0.07 dex) and Mg, for which the mean difference is larger and around 0.3 dex. 
There is range in the values for the standard deviation of the mean. For Si and Mn, $\sigma$ is $\leq$ 0.10 dex, while the Mg and Al abundances display a higher and similar level of scatter ($\sigma$= 0.17 dex). 

One element to discuss in particular is iron, as its abundance is typically used as an indicator of stellar metallicity. Abundances are often studied relative to metallicity and adopt the Sun as a reference, or use the [X/Fe] notation, where X represents the elemental abundance of each species. The derived iron abundances [X/H] are on average 0.24 dex higher than those from the DR16. (A similar abundance offset was also found in \citealt{Souto2021}, who analyzed a sample of M dwarfs belonging to the Coma Berenices open cluster.) In this context, we expect the [X/Fe] ratios between this work and DR16 to be more similar, as the offset is about $\sim$ 0.21 dex for most of the studied elements, removing some of the systematic differences when examining the variation with respect to the metallicity. 
A future effort using the methodology in this study will produce a revised set of improved results for all M dwarfs observed by APOGEE.

\begin{figure*}
\begin{center}
\includegraphics[angle=0,width=0.49\linewidth,clip]{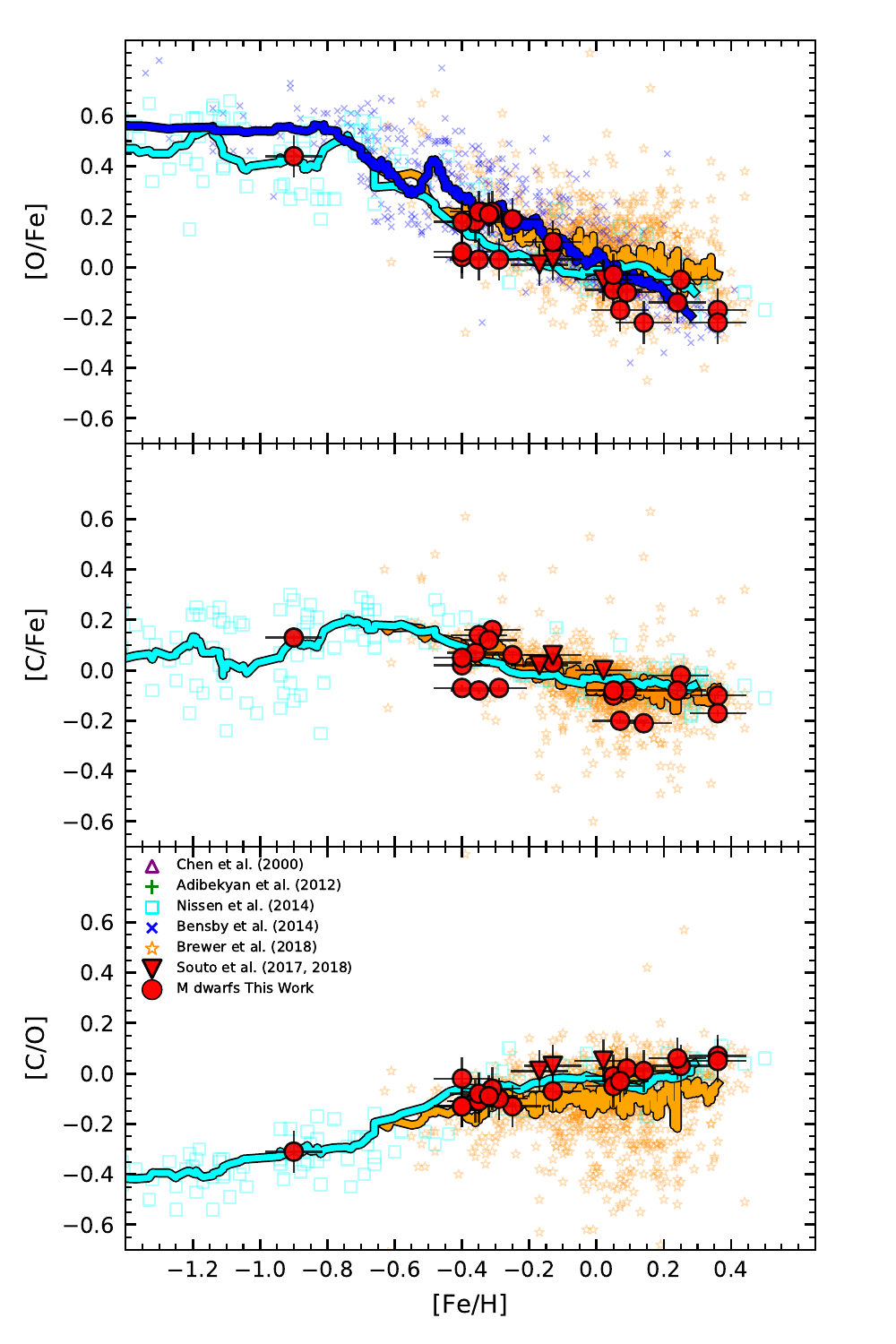}
\includegraphics[angle=0,width=0.49\linewidth,clip]{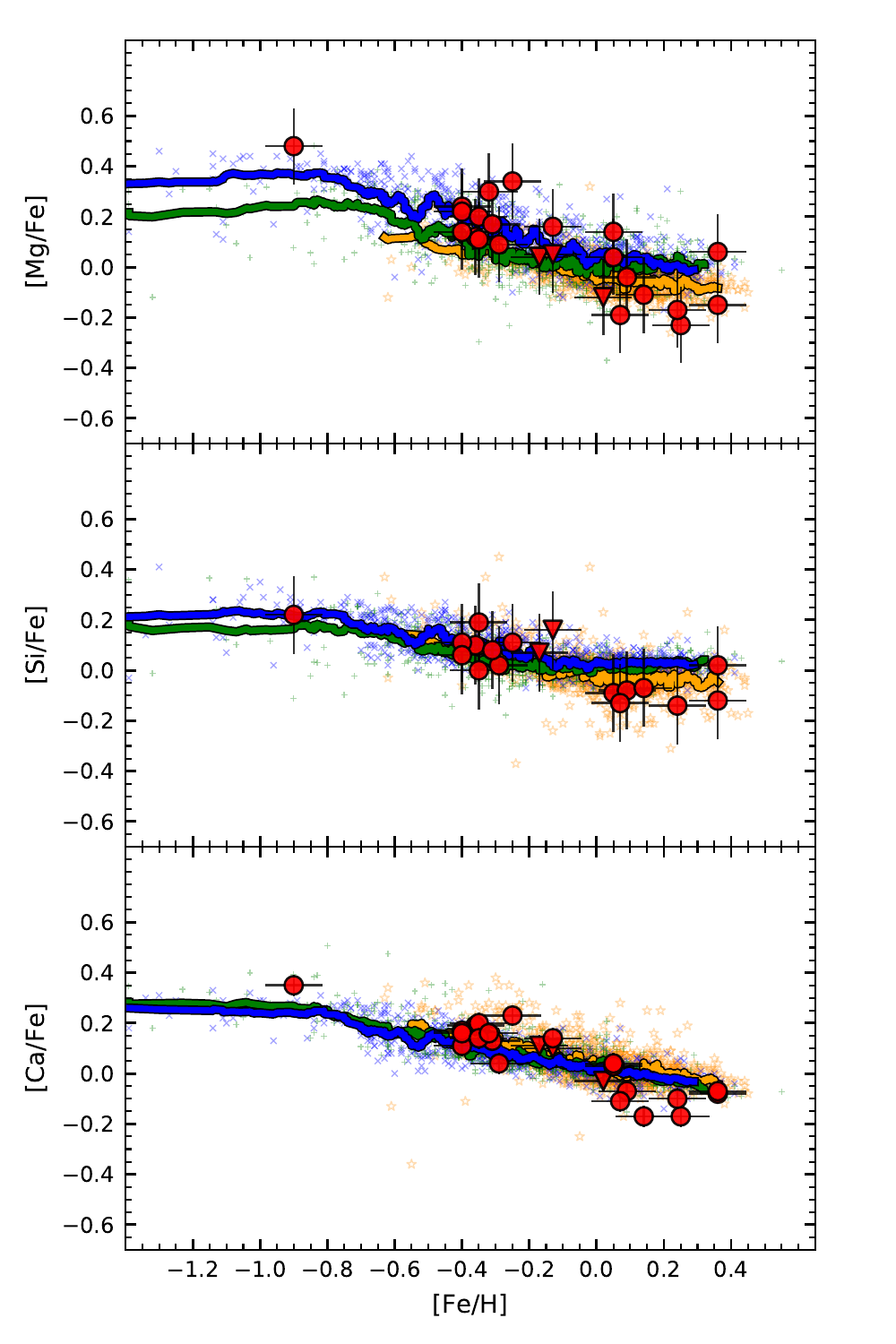}
\caption{Left panel: the Galactic trends of [O, C, and C/O over Fe] as a function of [Fe/H] for the studied M dwarfs (red symbols). The Galactic field star results are from \cite{Nissen2014} (cyan open squares), \cite{Bensby2014} (blue Xs), and \cite{BrewerFischer2018} (orange stars). Right panel: The Galactic trends of the $\alpha$ elements [Mg, Si, and Ca/Fe] as a function of [Fe/H] for the studied M dwarfs (red symbols). The Galactic field star results are from  \cite{Adibekyan2012} (green pluses), \cite{Bensby2014} (blue Xs), \cite{BrewerFischer2018} (orange stars).}

\end{center}
\label{CO_odd_abu}
\end{figure*}

\begin{figure*}
\begin{center}
\includegraphics[angle=0,width=0.49\linewidth,clip]{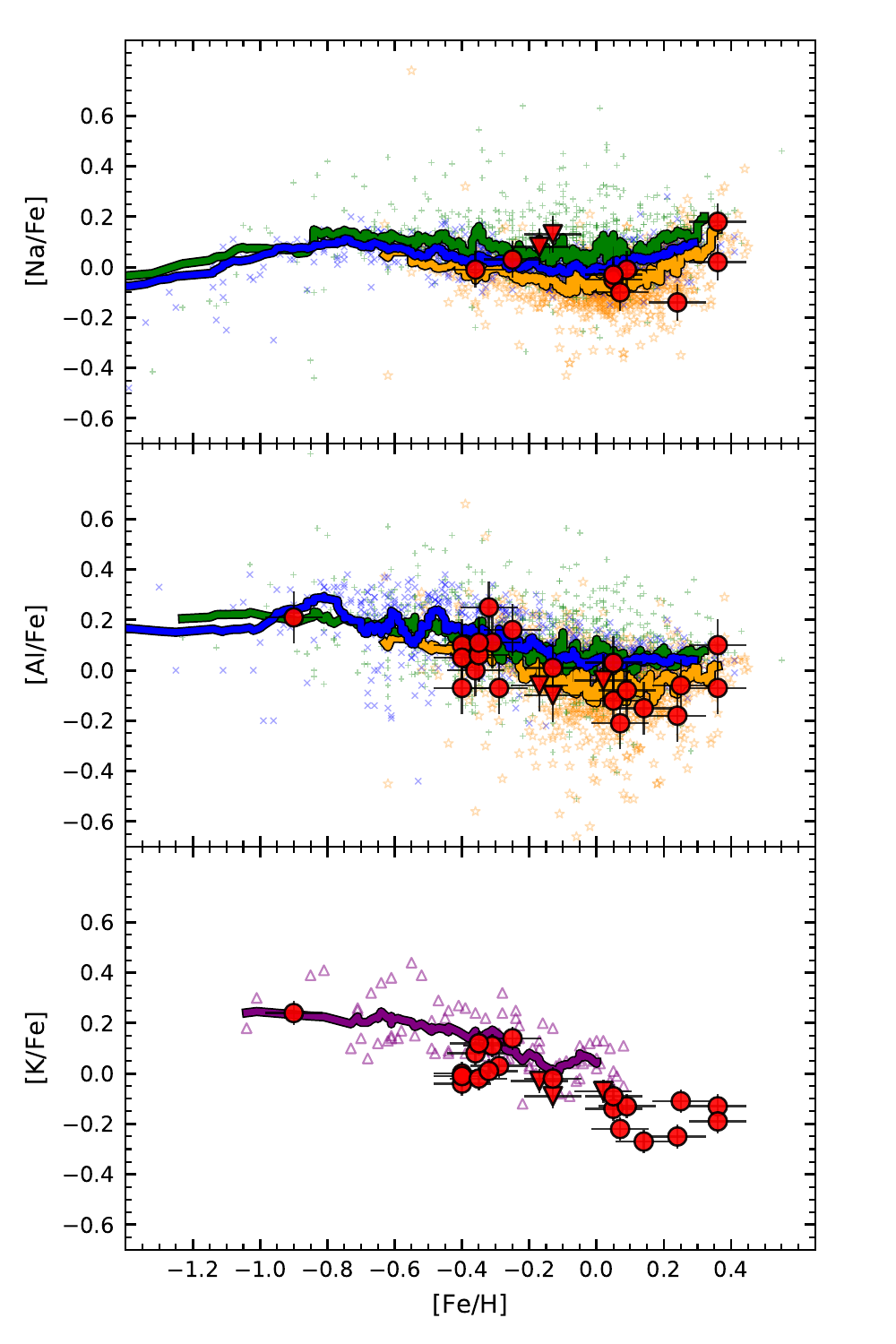}
\includegraphics[angle=0,width=0.49\linewidth,clip]{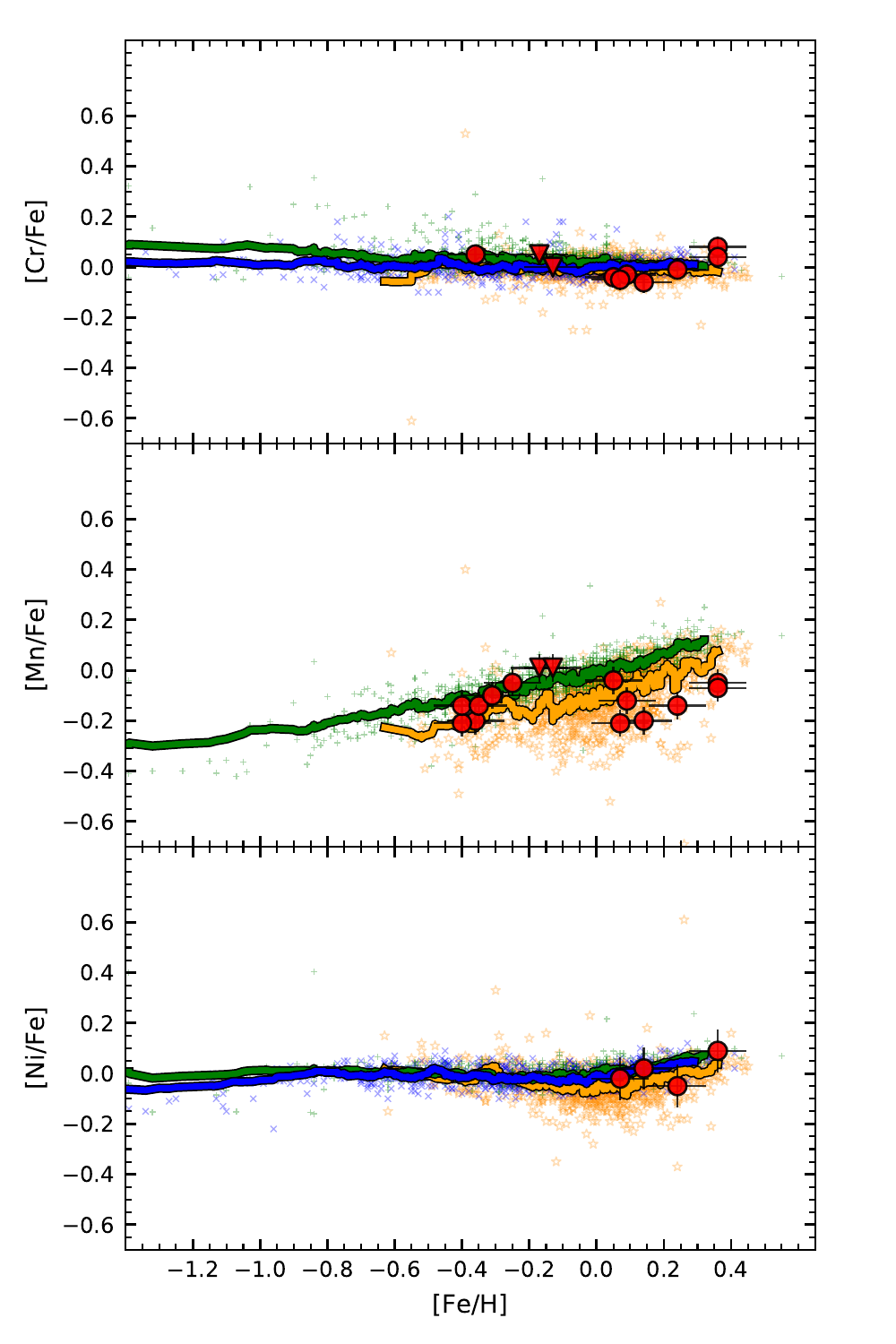}
\caption{Left panel: the Galactic trends of [Na, Al, and K/Fe] as a function of [Fe/H] for the studied M dwarfs (red symbols). The Galactic field star results are from  \cite{Adibekyan2012} (green pluses), \cite{Bensby2014} (blue Xs), \cite{Chen2000} (purple triangles), \cite{BrewerFischer2018} (orange stars). Right panel: the Galactic trends of the iron-peak elements [Cr, Mn, and Ni/Fe] as a function of [Fe/H] for the studied M dwarfs (red symbols). The Galactic field star results are from  \cite{Adibekyan2012} (green pluses), \cite{Bensby2014} (blue Xs), \cite{BrewerFischer2018} (orange stars).}
\end{center}
\label{alpha_iron_abu}
\end{figure*}

\subsection{The Galactic Abundance Trends}

The targeted M dwarfs are all within 100 pc and constitute a sample of low-mass stars in the Solar neighborhood. The abundance results for eleven chemical elements for these M dwarfs offer the possibility, for the first time in the literature, to begin an investigation of chemical evolution trends via canonical diagrams of [X/Fe] versus [Fe/H] that include these low-mass M dwarfs -- the dominant number population in the Galaxy.
Such diagrams are presented in Figures \ref{CO_odd_abu} (for C, O, Mg, Si and Ca) and \ref{alpha_iron_abu} (for Na, Al, K, Cr, Mn and Ni), where the abundance results for the M dwarfs are shown as red circles, with abundance uncertainties displayed for each star; results for the exoplanet-hosting M dwarfs Kepler 186, Kepler 138, and Ross 128 from our previous studies \cite{Souto2017,Souto2018Ross} are also displayed (red triangles).
The [X/Fe] versus [Fe/H] trends for the M dwarfs shown in the different panels of Figures \ref{CO_odd_abu} and \ref{alpha_iron_abu} are consistent with what is expected from Galactic chemical evolution and the behavior observed in field stars of spectral types FGK. Included in these figures are also comparisons with abundances from the literature. 

Results for the [O/Fe], [C/Fe], [C/O], [Mg/Fe], [Si/Fe], and [Ca/Fe] ratios are shown in the different panels of Figure \ref{CO_odd_abu}. The solid lines in the figure correspond to the running mean of 20 periods of the abundance results from the reference studies.
Focusing first on oxygen, the various literature results delineate a trend that is expected from Galactic chemical evolution, as they map the decreasing [O/Fe] ratio as SN Ia begin to contribute with iron. 
We make no distinction between the low- and high- $\alpha$ sequences found for the Milky Way disk \citep{Nidever2014}. Overall, the behavior of [O/Fe] versus [Fe/H] is similar for all studies in the regime where they overlap in metallicity; our M dwarf abundance ratios also follow a similar trend.
A closer examination, however, reveals some systematic differences between the sets of results. It is apparent that our M dwarf abundances are in better agreement with the median abundance relation from the results in \cite{Nissen2014} (cyan squares), who studied carbon and oxygen abundances in a sample of FG main-sequence stars in the solar neighborhood from the C I line at $\lambda$5052.5\AA, and O I from both the triplet lines at $\lambda$7777\AA\ and the forbidden line at $\lambda$6300\AA. In addition, the M dwarf in our sample with the lowest metallicity ([Fe/H] $\sim$ -0.90) has significantly enhanced in [O/Fe], as expected from the production of oxygen in SN II, and it follows the [O/Fe] distribution for metal-poor stars ([Fe/H]$<$-0.8) in \cite{Nissen2014}. 
\cite{Bensby2014} also analyzed the O I triplet lines at $\lambda$7777\AA\ in a sample of 714 F and G dwarf and subgiant stars. Our M dwarf results lie roughly 0.10 dex below their median abundance curve (blue curve). 
The [O/Fe] vs [Fe/H] trend obtained from the large sample of FGK Kepler planet host stars in \cite{BrewerFischer2018} (orange curve) appear to be flatter compared to the other studies, including the results for the M dwarfs. There is also a clearly larger scatter in their oxygen abundance results (orange symbols), spreading from [O/Fe] $\sim$ -0.40 up to +0.90 dex.

For carbon, there is a modest enhancement in [C/Fe] as [Fe/H] decreases; this ratio stays roughly constant for lower metallicities. Carbon is formed both in low-mass and high-mass stars. The results from all studies (this one included) agree quite well, without significant systematic differences. Again, the derived C abundances for the M dwarfs (red circles) agree with the FG dwarf results from \cite{Nissen2014}. 
Our carbon results show slightly more scatter than theirs, but the match is nonetheless quite good ($<$ 0.10 dex across the [Fe/H] vs [O/Fe] diagram). The lowest metallicity M dwarf in our sample is also in line with [C/Fe] ratios for stars with similar metallicities, falling roughly in the median of the distribution of \cite{Nissen2014}. 
The M dwarf [C/Fe] vs [Fe/H] results are also consistent with those from \cite{BrewerFischer2018} (orange symbols/curve), in this case also for FG stars, although there are several stars in their sample having significantly lower and higher values for [C/Fe] than ours.

Examining the [C/O] abundance ratio versus [Fe/H] diagram (left bottom panel of Figure \ref{CO_odd_abu}), reveals that the M dwarf [C/O] ratios decrease with decreasing metallicity with a small scatter ($\sim$ -0.20 -- + 0.00 dex), again showing excellent agreement with such ratios in \cite{Nissen2014}.

Results for the $\alpha$ elements Mg, Si, and Ca are shown, respectively, in the right top, middle and bottom panels of Figure \ref{CO_odd_abu}.
The overall behavior of Mg/Fe, Si/Fe, and Ca/Fe with metallicity is similar to what is seen for oxygen-over-iron and what is expected from Galactic chemical evolution; the [Mg, Si, Ca /Fe] ratios exhibit an increase as [Fe/H] decreases, while a flat and enhanced [$\alpha$/Fe], corresponding to early production in SN II, is reached near [Fe/H] $\sim$ -1. 
An additional comparison for Mg, Si and Ca in the Galactic disk was performed by \cite{Adibekyan2012}, who analyzed a large sample of 1111 FGK dwarf stars from the HARPS planet search program. The Mg, Si and Ca abundances obtained here for the M dwarfs agree with the literature results, although our [Mg/Fe] ratios show more scatter than the other studies. In addition, for metallicities solar and above, the M dwarf [$\alpha$/Fe] values are systematically lower than the literature Si and Ca abundances.
Similarly to what is found for oxygen, the [Mg/Fe], [Si/Fe], and [Ca/Fe] ratios in the lowest metallicity studied M dwarf are elevated; for Si there is good agreement with the median abundances from the sample in both \cite{Adibekyan2012} and \cite{Bensby2014}, while we find a larger enhancement in [Mg/Fe] and [Ca/Fe] than in \cite{Adibekyan2012} and \cite{Bensby2014}.

In the left panels of Figure \ref{alpha_iron_abu}, we present abundance results for the odd-Z elements: Na, Al, and K. There are many fewer M dwarfs with measurable Na abundances when compared to the other elements studied here, due to the fact that the Na I lines in the APOGEE M dwarf spectra are weak and not measurable for all stars in the sample.
The [Na/Fe] ratios obtained for the M dwarfs are within the Galactic range when considering the literature values shown in the figure, although some systematic differences between the studies are present. For both Na and Al the sequences defined by the median abundance values for the sample in each study are similar, but there are systematic differences, which are not surprising, at the level of 0.1 - 0.2 dex.
The [Na/Fe] ratios from \cite{Bensby2014} overlap with the higher envelope in [Na/Fe] from \cite{BrewerFischer2018}, but the latter displays a lower sequence in [Na/Fe] that is not present in \cite{Bensby2014} that reduces the median abundance values.
\cite{Adibekyan2012} [Na/Fe] ratios tend to scatter to higher values when compared to the other results. 
Our sodium abundances for the M dwarfs fall mostly along a low [Na/Fe] sequence and are in better agreement with the median values in \cite{BrewerFischer2018}; the Galactic trend of [Na/Fe] as a function of [Fe/H] displays the typical ``peanut shape'', with a clear upturn in [Na/Fe] for higher metallicities. 
The [Al/Fe] versus [Fe/H] abundance pattern obtained for the studied M dwarfs (Figure \ref{alpha_iron_abu}) is reminiscent of that of Na; here again our [Al/Fe] results are overall lower than \cite{Bensby2014} and \cite{Adibekyan2012} and are in better agreement with  \cite{BrewerFischer2018}. Our results also show an upturn in [Al/Fe] for metallicities roughly above solar.
For potassium there are fewer studies available in the literature, but we can compare our results for the M dwarfs with those from \cite{Chen2000}. Their median [K/Fe] abundances rise as the metallicity decreases for metallicities solar and below (bottom left panel of Figure \ref{alpha_iron_abu}). The [K/Fe] abundance ratios derived for the M dwarfs follow a similar sequence, but are offset by roughly -0.15 dex. At solar metallicity, [Fe/H = 0], the [K/Fe] ratio for M dwarfs is also below solar by roughly 0.2 dex (although the \citealt{Chen2000} sample do not populate this parameter space). 

For the iron peak elements Cr, Mn, and Ni (shown in the right panels of Figure \ref{alpha_iron_abu}), the derived M dwarf abundances follow the same scale as Fe. 
The [Cr/Fe] and [Ni/Fe] ratios display similar behaviors, and there is good agreement between the results in all studies, including the Cr and Ni abundances for the M dwarfs derived here. (Ni abundances have not been derived for M dwarfs with metallicities roughly below solar because the Ni I spectral lines become too weak for measurement measurable, and there is only one reliably measurable Cr I line in the APOGEE spectra.) 
The Mn results display an overall larger scatter when compared to Cr and Ni. There is flat trend similar to those of Cr and Ni, although with a larger scatter. In fact, the Mn abundances derived in this work follow the trend of Mn abundances derived from non-LTE, as seen in \cite{Bergemann2008Mn} for example, suggesting that non-LTE deviations are small for the lines in the APOGEE NIR spectra of mid-early M dwarf stars.

The recent work of \cite{Ishikawa2021} studied M dwarfs [X/Fe] ratios as a function of [Fe/H] for five species in common with this work. The overall trends of [Mg/Fe], [Cr/Fe], and [Mn/Fe] vs. [Fe/H] in \cite{Ishikawa2021} are similar to the ones presented here, while their [Na/Fe] values vs. [Fe/H] fall somewhat lower by $\sim$0.1 dex, and their [Ca/Fe] vs. [Fe/H] trend is flatter.

\section{Conclusions}

M dwarfs are the most abundant stellar class in the Galaxy and yet their chemical content has just recently begun to be probed in detail. We report on abundances of the elements C, O, Na, Mg, Al, Si, K, Ca, Cr, Mn, Fe, and Ni measured in a sample of nearby field M dwarfs observed by the high-resolution NIR APOGEE survey. Our sample contains eleven M dwarf stars which are secondaries in wide binary systems, plus ten M dwarfs with measured interferometric radii. The latter are benchmark comparisons for the effective temperatures, and in principle, the wide binaries offer the possibility to probe if there are large and spurious systematic differences between detailed chemical abundances of M dwarfs in comparison to the warmer primary stars analyzed from high-resolution optical spectra in the literature. 

The use of warm primaries in binary systems as benchmark comparisons for M dwarfs makes the crucial assumption that the chemical compositions of the M dwarfs and the primary stars are the same. There may be some level of diffusion effects, which depend on the age and mass of the warm primaries; such effects reduce the original abundances by typically less than 0.10 dex, at most for turn off stars and minimally for the M dwarfs.  Diffusion effects can, however, be minimized by investigating abundance ratios, for example, relative to iron, as all chemical elements suffer from roughly similar levels of diffusion within a few hundredths of dex \citep{Choi2016}. The comparison of the [C/Fe], [O/Fe], [Na/Fe], [Mg/Fe], [Al/Fe], [Si/Fe], [Ca/Fe], [Mn/Fe], and [Ni/Fe] abundance ratios for the M dwarfs and their warm primaries gives an offset of 0.01, -0.05, -0.05, 0.01, -0.08, -0.06, -0.08, 0.05, -0.04, respectively (with a mean abundance difference of -0.04 $\pm$ 0.04 dex). 
This indicates that the abundance scale for the M dwarfs in this study is precise, within the abundance uncertainties, and that the M dwarf metallicities and detailed abundance distributions are close matches to their hotter primaries. 

The chemical abundances of the M dwarfs were derived using a different methodology from that adopted by the APOGEE automatic abundance pipeline ASPCAP \citep{GarciaPerez2016}, although we used the same APOGEE updated line list \citep{Smith2021}. 
A comparison of our abundance results with those from ASPCAP DR16 indicates significant systematic offsets for all studied elements.


This sample of cool stars lies in the Solar neighborhood, having distances within 100 pc, and represents one of the first samples of M dwarfs to have the abundances of twelve chemical elements, along with chemical evolution trends investigated. 
Overall, the behavior of [X/Fe]  versus  [Fe/H] values, as shown in Figures  \ref{CO_odd_abu}  and  \ref{alpha_iron_abu}, are within what is expected from Galactic chemical evolution, as well as in agreement with that observed for field FGK stars from the literature. 

Beginning with carbon, there is a modest enhancement in [C/Fe] as [Fe/H] decreases and this ratio stays roughly constant for lower metallicity stars; the [C/O] ratios decrease with decreasing metallicity with small scatter ($\sim$ -0.20 -- + 0.00 dex). The results of [C/O], in particular, compare well with those of \cite{Nissen2014} and \cite{BrewerFischer2018}.
The overall behavior of the $\alpha$ elements O, Mg, Si, and Ca with metallicity is in line with what is expected from Galactic chemical evolution. 
The [O/Fe], [Mg/Fe], [Si/Fe], and [Ca/Fe] ratios rise as the metallicity declines, while the lowest metallicity M dwarfs have enhanced $\alpha$/Fe: Si and Ca measurements display good agreement with the median abundances from \cite{Adibekyan2012} and \cite{Bensby2014}, while for Mg there is also good agreement but our results show more scatter.

The abundance trends obtained for odd-Z elements Na and Al obtained for the M dwarfs are within the range of the literature values. At solar metallicities and above the M dwarfs exhibit the same trend of increasing Na/Fe and Al/Fe values as found for the FGK stars.
The [K/Fe] abundances rise as the metallicity decreases for metallicities solar and below.
For the iron peak elements Cr, Mn, and Ni, the derived M dwarf abundances overall track that of iron closely. The Mn abundances derived follow the trend from the non-LTE calculations from \citep{Bergemann2008Mn}, suggesting that non-LTE deviations are small for the Mn I lines in the APOGEE spectra of mid-early M dwarf stars.

The methodology in this work and previous works of our team (\citealt{Souto2020, Souto2021}) will be the basis of the reanalysis of all M dwarfs observed by APOGEE in future work. Also, the chemical abundances for the three hosting exoplanet M dwarfs in our sample can be used to study planet-stars connection and interior structure of those exoplanets.

\acknowledgments

KC and VS acknowledge that their work here is supported, in part, by the National Science Foundation through NSF grant AST-2009507, as well as by the National Aeronautics and Space Administration under Grant 16-XRP16\_2-0004, issued through the Astrophysics Division of the Science Mission Directorate. 
DAGH, OZ, and TM acknowledge support from the State Research Agency (AEI) of the Ministry of Science, Innovation and Universities (MCIU) and the European Regional Development Fund (FEDER) under grant AYA2017-88254-P.
SM acknowledge the Penn State's Center for Exoplanets and Habitable worlds.
B.R-A. acknowledges funding support from FONDECYT through grant 11181295.

Funding for the Sloan Digital Sky Survey IV has been provided by the Alfred P. Sloan Foundation, the U.S. Department of Energy Office of Science, and the Participating Institutions. SDSS-IV acknowledges
support and resources from the Center for High-Performance Computing at the University of Utah. The SDSS web site is www.sdss.org.

SDSS-IV is managed by the Astrophysical Research consortium for the 
Participating Institutions of the SDSS Collaboration including the 
Brazilian Participation Group, the Carnegie Institution for Science, 
Carnegie Mellon University, the Chilean Participation Group, the French Participation Group, Harvard-Smithsonian Center for Astrophysics, 
Instituto de Astrof\'isica de Canarias, The Johns Hopkins University, 
Kavli Institute for the Physics and Mathematics of the Universe (IPMU) /  
University of Tokyo, Lawrence Berkeley National Laboratory, 
Leibniz Institut f\"ur Astrophysik Potsdam (AIP),  
Max-Planck-Institut f\"ur Astronomie (MPIA Heidelberg), 
Max-Planck-Institut f\"ur Astrophysik (MPA Garching), 
Max-Planck-Institut f\"ur Extraterrestrische Physik (MPE), 
National Astronomical Observatory of China, New Mexico State University, 
New York University, University of Notre Dame, 
Observat\'orio Nacional / MCTI, The Ohio State University, 
Pennsylvania State University, Shanghai Astronomical Observatory, 
United Kingdom Participation Group,
Universidad Nacional Aut\'onoma de M\'exico, University of Arizona, 
University of Colorado Boulder, University of Oxford, University of Portsmouth, 
University of Utah, University of Virginia, University of Washington, University of Wisconsin, 
Vanderbilt University, and Yale University.

\facility {Sloan}

\software{Turbospectrum (\citealt{AlvarezPLez1998}, \citealt{Plez2012}), MARCS (\citealt{Gustafsson2008}), Bacchus (\citealt{Masseron2016}), Matplotlib (\citealt{matplotlib}), Numpy (\citealt{numpy}).}


\begin{thebibliography}{}
\bibitem[Abolfathi et al.(2018)]{DR14} Abolfathi, B., Aguado, D.~S., Aguilar, G., et al.\ 2018, \apjs, 235, 42
\bibitem[Adibekyan et al.(2012)]{Adibekyan2012} Adibekyan, V.~Z., Sousa, S.~G., Santos, N.~C., et al.\ 2012, \aap, 545, A32
\bibitem[Ahn et al.(2014)]{DR10} Ahn, C.~P., Alexandroff, R., Allende Prieto, C., et al.\ 2014, \apjs, 211, 17
\bibitem[Ahumada et al.(2020)]{DR16} Ahumada, R., Prieto, C.~A., Almeida, A., et al.\ 2020, \apjs, 249, 3. doi:10.3847/1538-4365/ab929e
\bibitem[Alam et al.(2015)]{DR12} Alam, S., Albareti, F.~D., Allende Prieto, C., et al.\ 2015, \apjs, 219, 12
\bibitem[Allard et al.(2000)]{Allard2000} Allard, F., Hauschildt, P.~H., \& Schwenke, D.\ 2000, \apj, 540, 1005
\bibitem[Alvarez \& Plez(1998)]{AlvarezPLez1998} Alvarez, R., \& Plez, B.\ 1998, \aap, 330, 1109
\bibitem[Barber et al.(2006)]{Barber2006} Barber, R.~J., Tennyson, J., Harris, G.~J., et al.\ 2006, \mnras, 368, 1087
\bibitem[Battistini \& Bensby(2015)]{Battistini2015} Battistini, C., \& Bensby, T.\ 2015, \aap, 577, A9
\bibitem[Bean et al.(2006)]{Bean2006} Bean, J.~L., Sneden, C., Hauschildt, P.~H., et al.\ 2006, \apj, 652, 1604
\bibitem[Bensby et al.(2014)]{Bensby2014} Bensby, T., Feltzing, S., \& Oey, M.~S.\ 2014, \aap, 562, A71
\bibitem[Bergemann \& Gehren(2008)]{Bergemann2008Mn} Bergemann, M., \& Gehren, T.\ 2008, \aap, 492, 823 
\bibitem[Bertelli Motta et al.(2018)]{BertelliMotta2018} Bertelli Motta, C., Pasquali, A., Richer, J., et al.\ 2018, \mnras, 478, 425. doi:10.1093/mnras/sty1011
\bibitem[Birky et al.(2020)]{Birky2020} Birky, J., Hogg, D.~W., Mann, A.~W., et al.\ 2020, \apj, 892, 31
\bibitem[Blanton et al.(2017)]{SDSS4} Blanton, M.~R., Bershady, M.~A., Abolfathi, B., et al.\ 2017, \aj, 154, 28
\bibitem[Bonfils et al.(2005)]{Bonfils2005} Bonfils, X., Delfosse, X., Udry, S., et al.\ 2005, \aap, 442, 635. doi:10.1051/0004-6361:20053046
\bibitem[Boyajian et al.(2012)]{Boyajian2012} Boyajian, T.~S., von Braun, K., van Belle, G., et al.\ 2012, \apj, 757, 112
\bibitem[Brewer \& Fischer(2018)]{BrewerFischer2018} Brewer, J.~M., \& Fischer, D.~A.\ 2018, \apjs, 237, 38
\bibitem[Carretta et al.(2013)]{Carretta2013} Carretta, E., Gratton, R.~G., Bragaglia, A., et al.\ 2013, \apj, 769, 40
\bibitem[Chavez \& Lambert(2009)]{ChavezLambert2009} Chavez, J., \& Lambert, D.~L.\ 2009, \apj, 699, 1906
\bibitem[Chen et al.(2000)]{Chen2000} Chen, Y.~Q., Nissen, P.~E., Zhao, G., et al.\ 2000, \aaps, 141, 491. doi:10.1051/aas:2000124
\bibitem[Choi et al.(2016)]{Choi2016} Choi, J., Dotter, A., Conroy, C., et al.\ 2016, \apj, 823, 102 
\bibitem[Covey et al.(2010)]{Covey2010} Covey, K.~R., Lada, C.~J., Rom{\'a}n-Z{\'u}{\~n}iga, C., et al.\ 2010, \apj, 722, 971
\bibitem[da Silva et al.(2015)]{daSilva2015_helio} da Silva, R., Milone, A. de C., \& Rocha-Pinto, H.~J.\ 2015, \aap, 580, A24
\bibitem[Delgado Mena et al.(2010)]{DelgadoMena2010} Delgado Mena, E., Israelian, G., Gonz{\'a}lez Hern{\'a}ndez, J.~I., et al.\ 2010, \apj, 725, 2349
\bibitem[Dotter et al.(2017)]{Dotter2017} Dotter, A., Conroy, C., Cargile, P., \& Asplund, M.\ 2017, \apj, 840, 99 
\bibitem[Gao et al.(2018)]{Gao2018} Gao, X., Lind, K., Amarsi, A.~M., et al.\ 2018, \mnras, 481, 2666. doi:10.1093/mnras/sty2414
\bibitem[Garc{\'\i}a P{\'e}rez et al.(2016)]{GarciaPerez2016} Garc{\'\i}a P{\'e}rez, A.~E., Allende Prieto, C., Holtzman, J.~A., et al.\ 2016, \aj, 151, 144
\bibitem[Gonz{\'a}lez-{\'A}lvarez et al.(2020)]{GonzalezAlvarez2020} Gonz{\'a}lez-{\'A}lvarez, E., Zapatero Osorio, M.~R., Caballero, J.~A., et al.\ 2020, \aap, 637, A93. doi:10.1051/0004-6361/201937050
\bibitem[Gunn et al.(2006)]{Gunn2006} Gunn, J.~E., Siegmund, W.~A., Mannery, E.~J., et al.\ 2006, \aj, 131, 2332
\bibitem[Gustafsson et al.(2008)]{Gustafsson2008} Gustafsson, B., Edvardsson, B., Eriksson, K., et al.\ 2008, \aap, 486, 951
\bibitem[Hargreaves et al.(2010)]{Hargreaves2010} Hargreaves, R.~J., Hinkle, K.~H., Bauschlicher, C.~W., et al.\ 2010, \aj, 140, 919
\bibitem[Henry et al.(2018)]{Henry2018} Henry, T.~J., Jao, W.-C., Winters, J.~G., et al.\ 2018, \aj, 155, 265
\bibitem[Hinkel et al.(2014)]{Hypatia} Hinkel, N.~R., Timmes, F.~X., Young, P.~A., et al.\ 2014, \aj, 148, 54
\bibitem[Holtzman et al.(2018)]{Holtzman2018} Holtzman, J.~A., Hasselquist, S., Shetrone, M., et al.\ 2018, \aj, 156, 125
\bibitem[Hunter(2007)]{matplotlib} Hunter, J.~D.\ 2007, Computing in Science and Engineering, 9, 90 
\bibitem[Ishikawa et al.(2020)]{Ishikawa2020} Ishikawa, H.~T., Aoki, W., Kotani, T., et al.\ 2020, \pasj, 72, 102. doi:10.1093/pasj/psaa101
\bibitem[Ishikawa et al.(2021)]{Ishikawa2021} Ishikawa, H.~T., Aoki, W., Hirano, T., et al.\ 2021, arXiv:2112.00173
\bibitem[J{\"o}nsson et al.(2018)]{Jonsson2018} J{\"o}nsson, H., Allende Prieto, C., Holtzman, J.~A., et al.\ 2018, \aj, 156, 126
\bibitem[J{\"o}nsson et al.(2020)]{Jonsson2020} J{\"o}nsson, H., Holtzman, J.~A., Allende Prieto, C., et al.\ 2020, \aj, 160, 120. doi:10.3847/1538-3881/aba592
\bibitem[Lindgren et al.(2016)]{Lindgren2016} Lindgren, S., Heiter, U., \& Seifahrt, A.\ 2016, \aap, 586, A100
\bibitem[Lindgren \& Heiter(2017)]{Lindgren2017} Lindgren, S., \& Heiter, U.\ 2017, \aap, 604, A97
\bibitem[Mahadevan et al.(2012)]{Mahadevan2012} Mahadevan, S., Ramsey, L., Bender, C., et al.\ 2012, \procspie, 8446, 84461S. doi:10.1117/12.926102
\bibitem[Majewski et al.(2017)]{Majewski2017} Majewski, S.~R., Schiavon, R.~P., Frinchaboy, P.~M., et al.\ 2017, \aj, 154, 94
\bibitem[Mann et al.(2013a)]{Mann2013_binarypaper} Mann, A.~W., Brewer, J.~M., Gaidos, E., et al.\ 2013, \aj, 145, 52
\bibitem[Mann et al.(2013b)]{Mann2013b} Mann, A.~W., Gaidos, E., \& Ansdell, M.\ 2013, \apj, 779, 188
\bibitem[Masseron et al.(2016)]{Masseron2016} Masseron, T., Merle, T., \& Hawkins, K.\ 2016, BACCHUS: Brussels Automatic Code for Characterizing High accUracy Spectra, ascl:1605.004
\bibitem[Miller \& Scalo(1979)]{Miller1979} Miller, G.~E., \& Scalo, J.~M.\ 1979, \apjs, 41, 513
\bibitem[Michaud et al.(2015)]{Michaud2015} Michaud, G., Alecian, G., \& Richer, J.\ 2015, Atomic Diffusion in Stars, Astronomy and Astrophysics Library, ISBN 978-3-319-19853-8. Springer International Publishing Switzerland, 2015.. doi:10.1007/978-3-319-19854-5
\bibitem[Mishenina et al.(2008)]{Mishenina2008} Mishenina, T.~V., Soubiran, C., Bienaym{\'e}, O., et al.\ 2008, \aap, 489, 923
\bibitem[Mishenina et al.(2013)]{Mishenina2013} Mishenina, T.~V., Pignatari, M., Korotin, S.~A., et al.\ 2013, \aap, 552, A128
\bibitem[Mishenina et al.(2015)]{Mishenina2015} Mishenina, T., Gorbaneva, T., Pignatari, M., et al.\ 2015, \mnras, 454, 1585
\bibitem[Montes et al.(2018)]{Montes2018} Montes, D., Gonz{\'a}lez-Peinado, R., Tabernero, H.~M., et al.\ 2018, \mnras, 479, 1332
\bibitem[Newton et al.(2014)]{Newton2014} Newton, E.~R., Charbonneau, D., Irwin, J., et al.\ 2014, \aj, 147, 20
\bibitem[Nidever et al.(2014)]{Nidever2014} Nidever, D.~L., Bovy, J., Bird, J.~C., et al.\ 2014, \apj, 796, 38. doi:10.1088/0004-637X/796/1/38
\bibitem[Nidever et al.(2015)]{Nidever2015} Nidever, D.~L., Holtzman, J.~A., Allende Prieto, C., et al.\ 2015, \aj, 150, 173
\bibitem[Nissen et al.(2014)]{Nissen2014} Nissen, P.~E., Chen, Y.~Q., Carigi, L., et al.\ 2014, \aap, 568, A25
\bibitem[Passegger et al.(2018)]{Passegger2018} Passegger, V.~M., Reiners, A., Jeffers, S.~V., et al.\ 2018, \aap, 615, A6
\bibitem[Pinsonneault et al.(2014)]{Pinsonneault2014} Pinsonneault, M.~H., Elsworth, Y., Epstein, C., et al.\ 2014, \apjs, 215, 19
\bibitem[Pinsonneault et al.(2018)]{Pinsonneault2018} Pinsonneault, M.~H., Elsworth, Y.~P., Tayar, J., et al.\ 2018, \apjs, 239, 32
\bibitem[Plez(2012)]{Plez2012} Plez, B.\ 2012, Turbospectrum: Code for spectral synthesis, ascl:1205.004
\bibitem[Quirrenbach et al.(2014)]{Carmenes2014} Quirrenbach, A., Amado, P.~J., Caballero, J.~A., et al.\ 2014, \procspie, 91471F
\bibitem[Pinamonti et al.(2018)]{Pinamonti2018} Pinamonti, M., Damasso, M., Marzari, F., et al.\ 2018, \aap, 617, A104. doi:10.1051/0004-6361/201732535
\bibitem[Pinsonneault et al.(1989)]{Pinsonneault1989} Pinsonneault, M.~H., Kawaler, S.~D., Sofia, S., et al.\ 1989, \apj, 338, 424. doi:10.1086/167210
\bibitem[Rajpurohit et al.(2018)]{Rajpurohit2018} Rajpurohit, A.~S., Allard, F., Teixeira, G.~D.~C., et al.\ 2018, \aap, 610, A19
\bibitem[Ram{\'\i}rez et al.(2012)]{Ramirez2012} Ram{\'\i}rez, I., Fish, J.~R., Lambert, D.~L., et al.\ 2012, \apj, 756, 46
\bibitem[Ram{\'\i}rez et al.(2013)]{Ramirez2013} Ram{\'\i}rez, I., Allende Prieto, C., \& Lambert, D.~L.\ 2013, \apj, 764, 78
\bibitem[Reddy et al.(2006)]{Reddy2006} Reddy, B.~E., Lambert, D.~L., \& Allende Prieto, C.\ 2006, \mnras, 367, 1329
\bibitem[Reiners et al.(2018)]{Reiners2018} Reiners, A., Zechmeister, M., Caballero, J.~A., et al.\ 2018, \aap, 612, A49
\bibitem[Rojas-Ayala et al.(2010)]{RojasAyala2010} Rojas-Ayala, B., Covey, K.~R., Muirhead, P.~S., et al.\ 2010, \apjl, 720, L113
\bibitem[Rojas-Ayala et al.(2012)]{RojasAyala2012} Rojas-Ayala, B., Covey, K.~R., Muirhead, P.~S., et al.\ 2012, \apj, 748, 93
\bibitem[Sarmento et al.(2021)]{Sarmento2021} Sarmento, P., Rojas-Ayala, B., Delgado Mena, E., et al.\ 2021, \aap, 649, A147. doi:10.1051/0004-6361/202039703
\bibitem[Schmidt et al.(2016)]{Schmidt2016} Schmidt, S.~J., Wagoner, E.~L., Johnson, J.~A., et al.\ 2016, \mnras, 460, 2611
\bibitem[Semenova et al.(2020)]{Semenova2020} Semenova, E., Bergemann, M., Deal, M., et al.\ 2020, \aap, 643, A164. doi:10.1051/0004-6361/202038833
\bibitem[Shetrone et al.(2015)]{Shetrone2015} Shetrone, M., Bizyaev, D., Lawler, J.~E., et al.\ 2015, \apjs, 221, 24
\bibitem[Shi et al.(2004)]{Shi2004} Shi, J.~R., Gehren, T., \& Zhao, G.\ 2004, \aap, 423, 683
\bibitem[Schuler et al.(2003)]{Schuler2003} Schuler, S.~C., King, J.~R., Fischer, D.~A., et al.\ 2003, \aj, 125, 2085. doi:10.1086/373927
\bibitem[Smith et al.(2021)]{Smith2021} Smith, V.~V., Bizyaev, D., Cunha, K., et al.\ 2021, \aj, 161, 254. doi:10.3847/1538-3881/abefdc
\bibitem[Skrutskie et al.(2006)]{2MASS} Skrutskie, M.~F., Cutri, R.~M., Stiening, R., et al.\ 2006, \aj, 131, 1163
\bibitem[Soubiran \& Girard(2005)]{Soubiran2005} Soubiran, C., \& Girard, P.\ 2005, \aap, 438, 139
\bibitem[Souto et al.(2016)]{Souto2016} Souto, D., Cunha, K., Smith, V., et al.\ 2016, \apj, 830, 35 
\bibitem[Souto et al.(2017)]{Souto2017} Souto, D., Cunha, K., Garc{\'\i}a-Hern{\'a}ndez, D.~A., et al.\ 2017, \apj, 835, 239
\bibitem[Souto et al.(2018)]{Souto2018} Souto, D., Cunha, K., Smith, V.~V., et al.\ 2018, \apj, 857, 14 
\bibitem[Souto et al.(2018)]{Souto2018Ross} Souto, D., Unterborn, C.~T., Smith, V.~V., et al.\ 2018, \apjl, 860, L15
\bibitem[Souto et al.(2019)]{Souto2019} Souto, D., Allende Prieto, C., Cunha, K., et al.\ 2019, \apj, 874, 97
\bibitem[Souto et al.(2020)]{Souto2020} Souto, D., Cunha, K., Smith, V.~V., et al.\ 2020, \apj, 890, 133
\bibitem[Souto et al.(2021)]{Souto2021} Souto, D., Cunha, K., \& Smith, V.~V.\ 2021, \apj, 917, 11. doi:10.3847/1538-4357/abfdb5
\bibitem[Stock et al.(2020)]{Stock2020} Stock, S., Nagel, E., Kemmer, J., et al.\ 2020, \aap, 643, A112. doi:10.1051/0004-6361/202038820
\bibitem[Su{\'a}rez-Andr{\'e}s et al.(2017)]{Suarez-Andres2017} Su{\'a}rez-Andr{\'e}s, L., Israelian, G., Gonz{\'a}lez Hern{\'a}ndez, J.~I., et al.\ 2017, \aap, 599, A96
\bibitem[Terrien et al.(2015)]{Terrien2015} Terrien, R.~C., Mahadevan, S., Deshpande, R., et al.\ 2015, \apjs, 220, 16
\bibitem[Tsuji \& Nakajima(2014)]{TsujiNakagima2014} Tsuji, T., \& Nakajima, T.\ 2014, \pasj, 66, 98
\bibitem[Tsuji et al.(2015)]{Tsuji2015} Tsuji, T., Nakajima, T., \& Takeda, Y.\ 2015, \pasj, 67, 26
\bibitem[Tsuji \& Nakajima(2016)]{TsujiNakagima2016} Tsuji, T., \& Nakajima, T.\ 2016, \pasj, 68, 13
\bibitem[van der Walt et al.(2011)]{numpy} van der Walt, S., Colbert, S.~C., \& Varoquaux, G.\ 2011, Computing in Science and Engineering, 13, 22
\bibitem[Veyette et al.(2017)]{Veyette2017} Veyette, M.~J., Muirhead, P.~S., Mann, A.~W., et al.\ 2017, \apj, 851, 26
\bibitem[Wilson et al.(2019)]{Wilson2019} Wilson, J.~C., Hearty, F.~R., Skrutskie, M.~F., et al.\ 2019, \pasp, 131, 055001. doi:10.1088/1538-3873/ab0075
\bibitem[Woolf \& Wallerstein(2005)]{WoolfWallerstein2005a} Woolf, V.~M., \& Wallerstein, G.\ 2005, \mnras, 356, 963
\bibitem[Woolf \& Wallerstein(2020)]{WoolfWallerstein2020} Woolf, V.~M. \& Wallerstein, G.\ 2020, \mnras, 494, 2718. doi:10.1093/mnras/staa878
\bibitem[Zamora et al.(2015)]{Zamora2015} Zamora, O., Garc{\'{\i}}a-Hern{\'a}ndez, D.~A., Allende Prieto, C., et al.\ 2015, \aj, 149, 181 
\end{thebibliography}
{}



\end{document}